# Ferroelectric or non-ferroelectric: why so many materials exhibit "ferroelectricity" on the nanoscale

Rama K. Vasudevan[1,2], Nina Balke[1,2], Peter Maksymovych[1,2],

Stephen Jesse[1,2] and Sergei V. Kalinin[1,2]

[1]Institute for Functional Imaging of Materials and [2]Center for Nanophase Materials Sciences,

Oak Ridge National Laboratory, Oak Ridge TN 37831, USA

**Abstract**

Ferroelectric materials have remained one of the foci of condensed matter physics and materials science for over 50 years. In the last 20 years, the development of voltage-modulated scanning probe microscopy techniques, exemplified by Piezoresponse force microscopy (PFM) and associated time- and voltage spectroscopies, opened a pathway to explore these materials on a single-digit nanometer level. Consequently, domain structures, walls and polarization dynamics can now be imaged in real space. More generally, PFM has allowed studying electromechanical coupling in a broad variety of materials ranging from ionics to biological systems. It can also be anticipated that the recent Nobel prize[1] in molecular electromechanical machines will result in rapid growth in interest in PFM as a method to probe their behavior on single device and device assembly levels. However, the broad introduction of PFM also resulted in a growing number of reports on nearly-ubiquitous presence of ferroelectric-like phenomena including remnant polar states and electromechanical hysteresis loops in materials which are non-ferroelectric in the bulk, or in cases where size effects are expected to suppress ferroelectricity. While in certain cases plausible physical mechanisms can be suggested, there is remarkable similarity in observed behaviors, irrespective of the materials system. In this review, we summarize the basic principles of PFM, briefly discuss the features of ferroelectric surfaces salient to PFM imaging and spectroscopy, and summarize existing reports on ferroelectric–like responses in non-classical ferroelectric materials. We further discuss possible mechanisms behind observed behaviors, and possible experimental strategies for their identification.

# I. INTRODUCTION

Ferroelectric materials remain one of the foci of materials science and condensed matter physics research, the interest stemming from both fundamental physics standpoints as well as for more practical purposes. From the applied perspective, this interest derives from their current use in a wide variety of active electromechanical and electrooptical elements in devices such as transducers, actuators, motors, sensors, non-volatile memories, etc.[2-7] Similarly, from the physics viewpoint, they provide a system to explore a gamut of functionalities ranging from the motion of elastic interfaces through disordered media and associated scaling laws for domain wall motion,[8-10] possibility of localized metal-insulator transitions[11-14] and order parameter coupling at ferroelectric domain walls,[15-18] statistical modeling of macroscopic polarization switching by microscopic switching units,[19-23] unique behaviors under spatial confinement,[24, 25] and many more. Thus, the study of ferroelectrics has remained pertinent even 90+ years since their initial discovery in a Rochelle salt, in 1921[26].

The advent of techniques to both synthesize and characterize thin-film ferroelectric materials in the early 1990s led to a burst of research in the area, which has continued to this day.[2] Key to this endeavor was the use of pulsed laser deposition, molecular beam epitaxy[27, 28] and other growth techniques to fabricate high quality epitaxial thin films, expanding the suite of available ferroelectrics by providing a pathway to explore materials that are often difficult to produce with phase purity in bulk forms. Furthermore, this provides complete control over thickness (down to the unit cell level[29-31]), allowing exploration of thickness effects on domain structure, polarization, piezoelectric coefficients, dielectric properties and more.[32] The availability of high quality single crystal substrates of varying lattice parameters also allowed investigation of strain effects in a controlled manner in ferroelectrics, later coined 'strain engineering,'[33] and which has numerous celebrated examples, with arguably the most visible being the discovery of multiferroic $BiFeO_3$,[34] as well as a strain-induced phase boundary at high levels of compressive stress[35] in the same system. The progress within the field of ferroelectrics is well captured through analysis of citation networks, using the CiteSpace program, and is illustrated in **Figure 1**. In Fig. 1(a), one can track the major keywords in papers relating to ferroelectrics, and observe their evolution in time. It is evident that research into thin film ferroelectrics has exploded in the last two decades, although very recently there has also been a push towards 'lead-free' ferroelectrics. This is further confirmed in Fig. 1(b), where individual highly cited papers are listed, and clustered according to the similarity of the terms contained within the papers. The clusters themselves are colored based on the average year of publication, and it is noticeable that many highly cited publications occurred since the early 2000s in this field, suggesting the continuous growth of the field.

The counterpart to the synthesis and exploration of the phase space of high quality epitaxial thin films by pulsed laser deposition was the necessity to image and manipulate ferroelectric domains in the nanoscale, which was met by the development of piezoresponse force microscopy.[36-40] In this review, we begin with a primer on the basics of piezoresponse force microscopy (PFM), as well as its extension to spectroscopy that allowed PFM to become the premier tool for nanoscale ferroelectric characterization in systems that were either too small or too electrically conducting for classical macroscopic Polarization-Electric Field (P-E) methods. We then discuss the spectroscopies that are commonly used with PFM, for probing time- and voltage dependent phenomena. We then proceed to explain in the third section that non-ferroelectric materials can produce PFM signals that are similar to those of standard ferroelectrics, with a table of the relevant observations from the literature. In the fourth section, we discuss observations of surface charge screening on ferroelectrics, as well as electrochemical strain microscopy and electrochemical writing. In the fifth section, we discuss various mechanisms involved, both physical and electrochemical, when voltage is applied to the atomic force microscope (AFM) tip, as well as their characteristics and frequency dispersions. In section VI, we discuss characterization methods beyond the traditional PFM methods that can differentiate some of these electrochemical processes from physical processes, with an overview of their opportunities and limitations. This directly leads to a discussion of the coupling between ionic/electronic and ferroelectric phenomena in section VII. We summarize in section VIII by noting the need for additional theory and techniques to provide more insight into the behavior of these materials in the high-field and high-stress situations afforded by the biased AFM tip in contact with the sample.

## II. PRINCIPLES OF PFM IMAGING AND SPECTROSCOPY

### II.A. BASIC PFM

The need for understanding of ferroelectric phenomena on the nanoscale has stimulated the development of local techniques for imaging and manipulation of ferroelectric materials, specifically probes sensitive to the local polarization orientation. The development of such probes can be roughly divided into the pre-1995 period and post-1995 period, as demarcated by the nearly simultaneous introduction of Piezoresponse Force Microscopy and (essentially similar) scanning probe microscopy methods by Kolosov and Gruverman,[41] Eng,[42] Takata,[40] Franke[43, 44] and later Khim.[45]

Prior to 1995, the methods for ferroelectric domain imaging were based on the then preponderant optical[46] and electron microscopy methods.[47-50] Indeed, the polarization dependence of optical and electro-optical properties of ferroelectrics enabled the broad use of optical microscopy for ferroelectric domain visualization.[51] Similarly, the domain-specific chemical reactivity of ferroelectrics enabled etching studies, in which topographic contrast is indicative of preexisting domain structure.[52] Finally, electron mirror polarization-dependent surface charge density of ferroelectrics was used as a basis for detection in several variants of scanning electron microscopy.[53] These techniques were generally limited to imaging at relatively low (~0.5 μm) resolution only, and local quantitative ferroelectric properties remained inaccessible. However, in conjunction with macroscopic electromechanical and P-E measurements,[46] these techniques formed the mainstay of ferroelectric characterization.

The situation changed drastically in the late 90s, as a result of the broad adoption of scanning probe microscopy (SPM) techniques by the physics, biology, and materials science communities.[54-58] It was recognized by several groups that the fundamental characteristics of ferroelectrics, namely electromechanical coupling[41, 59], polarization dependent surface charge density,[43, 44, 60-62] polarization dependent surface chemistry and friction,[63-69] or optical properties[70] could be employed as a basis for detection in SPM. A number of authors reported the difference in friction properties between dissimilar ferroelectric domains, enabling high (sub-10 nm) resolution of domain structures of materials such as GASH (guanidinium aluminum sulfate hexahydrate), TGS (triglycine sulfate), etc. in friction force microscopy.[63-69] However, the nature and quantitative mechanisms of the observed signal remained highly uncertain. In parallel, several groups explored ferroelectric domain imaging using Kelvin Probe Force Microscopy (KPFM) and similar techniques,[29, 30, 37-39, 71-73] which are variants of non-contact SPM techniques sensitive to long-range electrostatic interactions. Early successes included visualization of domain structures, with the contrast attributed directly to the stray electrostatic field above the surface due to (unscreened) polarization bound charges. However, a set of studies by Kalinin and Bonnell demonstrated that this naïve interpretation is incorrect. Namely, quantitative analysis of the measured potential distributions including field strength and spatial behavior demonstrated that it is consistent with a dipole layer, rather than a charge sheet model.[74] Similarly, variable temperature and time dependent studies enabled observation of transient potential increase and retention above the transition temperature,[60, 61, 75, 76] temperature induced potential inversion and isopotential point observation,[77] as well as charge shadows after moving domain walls.[74] Most strikingly, the observed domain potential was shown to be opposite to that of polarization charge, reminiscent of charge inversion in colloidal systems. These observations were explained by the model of polarization almost completely screened by surface ionic charges, where the ionic charge dynamics are very slow (hours) compared to ferroelectric behavior. It was further shown that thermodynamic and kinetic parameters of this process can be inferred from KPFM data. Since then, this chemical screening behavior was explored in depth in a beautiful set of experimental and theoretical work by a group at Argonne,[78, 79] and further evidence of surface screening behavior was gained via observations of e.g. back switching[80-84] and chaotic screening of ferroelectrics by PFM.[85, 86] However, these studies demonstrated that KPFM (or any other long-range field contrast) is controlled by the interplay between polarization and (poorly understood and controlled) screening and charging processes, and does not provide information on ferroelectric behavior *per se*. Furthermore, the spatial resolution of KPFM is limited to ~50 nm, above the range of interest for ferroelectric materials often containing nanometer-scale domain structures.

Finally, the largest body of SPM work in ferroelectrics was generated by contact voltage-modulated techniques (See **Figure 2** for schematic). In this case, the SPM tip in a (strong) mechanical contact with the surface is biased by an oscillating voltage, and the electromechanical response of the surface is detected as a first harmonic of the cantilever deflection, $D_{ac}$. This principle was used to demonstrate high-resolution imaging of ferroelectric domain structures, often with sub 100 nm resolution. Furthermore, it was shown that application of bias pulses $V_{dc}$ can be used to switch the ferroelectric state of the surface, essentially enabling domain writing. This approach was immediately explored as a potential high-density non-volatile storage (ferroelectric hard drive), and many PFM practitioners of late 90s spent parts of their careers in the data storage industry pursuing this direction (the direction later overtaken by the emergence of phase change[87, 88] and electroresistive memories in 2005-2006[89-91]). However, these early studies differed in the interpretation of observed signal – namely, inverse piezoelectric effect (hence the name Piezoresponse Force Microscopy coined by Gruverman[92]) vs. contact electrostatic forces.

The relative magnitude of these phenomena were explored by Kalinin and Bonnell,[93, 94] giving rise to the concept of weak and strong indentation regimes, with preponderant electromechanical coupling in typical PFM operation conditions for strong ferroelectrics. The non-local electrostatic forces were studied by Harnagea *et al.*,[95] who explored the frequency dependence of the piezoelectric and electrostatic contributions to the PFM signal, in addition to the effect of the loading force. Further work on exploring the signal formation mechanisms, resolution theory, cross-talk and environmental effects came from Kalinin and Jesse.[96-98]

For PFM of a ferroelectric material, one can consider that the (local part of) mechanical displacement $D$ measured will be equal (as a first approximation) to[99]

$$D = D_{PE} + D_{ES} = d_{eff}V_{ac} + k^{*-1}C'(V_{dc} - V_{SP})V_{ac} \qquad (1)$$

where $D_{PE}$ is the piezoelectric contribution, $D_{ES}$ is the electrostatic contribution, $d_{eff}$ is the effective converse piezoelectric coefficient, $k^*$ is the contact stiffness, $C'$ is the capacitance gradient between tip and sample, $V_{SP}$ is the surface potential, $V_{dc}$ is the DC voltage applied to the tip, and $V_{ac}$ is the probing AC voltage (also applied to the tip). Typically, as a first order approximation $d_{eff}$ is invariant on $V_{dc}$ if no domain switching takes place, and therefore, the displacement from the piezoelectric contribution is independent of DC voltage (for a uniformly poled sample or domain). However, the electrostatic contribution scales with the contact potential difference, and thus, is expected to be linear.

In the 15 years since, the electromechanical model has become the leading paradigm for the observed contrast. In particular, it was shown that in many cases the quantified responses are close to the longitudinal piezoelectric coefficient $d_{33}$ of material, i.e. response expected for material in the uniform field. A number of analytical and numerical studies of the PFM mechanism were reported using exact solutions for a piezoelectric indentation model,[100-107] a numerical decoupled model,[108] and an analytical decoupled model.[109-115] Similar approaches were pursued by other groups.[116-120] These analyses were extended to quantitatively describe a ferroelectric domain wall profile in vertical and lateral PFM[115, 121, 122], orientational dependence of the PFM signal,[112, 123] and to quantify piezoresponse spectroscopy[124], as will be discussed below. Perhaps even more importantly, the PFM signal (more rigorously, electromechanical surface displacement) was shown to be independent of contact area, ensuring that PFM, unlike force based SPMs, can be an intrinsically quantitative technique. Overall, the ease of use and quantitative nature (stemming from a purely electromechanical interpretation) of PFM has laid the foundation for its broad adoption by ferroelectrics community.[125, 126]

The early studies of PFM focused on both static imaging of domain structures, as well as basic tip-induced domain writing experiments for ferroelectric crystals, thin films and other nanostructures. This included high-resolution imaging of ferroelectric and ferroelastic domains in $PbZr_xTi_{1-x}O_3$ (PZT) (and associated asymmetry in the vertical PFM profile across these domains[127]), patterning of domains in nanostructures of PZT, $SrBi_2Ta_2O_9$ (SBT) and $BaTiO_3$ (BTO) for their potential use as memory cells[128], vertical and lateral domain imaging and writing in single crystal TGS and BTO,[129] and domain lithography.[92] It was quickly realized that the technique would also be useful to study polarization relaxation and domain wall motion, which were explored by multiple groups worldwide.[130-133] For example, polarization reversal in written domains in PZT thin films demonstrated the role of the ferroelastic domain wall as a pinning site, directly displayed domain faceting, as well as lateral domain switching during 180° polarization reversal.[127, 134] Further studies extended the concept by directly using PFM data to investigate the regimes of domain wall motion (creep)[135], and quantify domain wall roughness.[136] Several examples where PFM is used for domain imaging and for exploring polarization reversal are shown in **Figure 3**.

Beyond ferroelectric materials, PFM was used to explore a number of systems with known piezoelectric behavior. For example, most biological systems ranging from wood to calcified and connective tissues are known to be piezoelectric[137-140], as a result of combination of optical activity and polar bonding in bio macromolecules. These behaviors often average out in macroscopic biological assemblies, since the piezoelectric effect is described by third rank tensor. However, on the nanoscale these polar regions are readily visualized by PFM, providing spectacular images of bones,[141] antlers[142] teeth,[143-147] and even butterfly wings.[147] Several examples are shown in **Figure 4**. One should of course be aware of the potential for topographic cross-talk in these systems, given the large topographic roughness. In some cases, the electromechanical activity was observed in symmetry-forbidden cases such as tobacco mosaic virus[148] and attributed to flexoelectric coupling.[149] Finally, several molecular crystals including glycine,[150] croconic acid,[151, 152] etc. were found to display ferroelectricity, providing new insights into the structure and properties of these materials and potentially leading to new applications. It can be anticipated that the recent Nobel prize in molecular electromechanical machines[1, 153, 154] will result in rapid growth in interest in PFM as a method to probe their behavior on single molecule and assembly levels.

### II.B. LOCAL SWITCHING SPECTROSCOPY

The interest in PFM grew further in the early 2000s, when several groups showed that the combination of PFM imaging and application of bias pulses enabled PFM spectroscopy.[155] In this, a set of bias pulses induces nucleation and growth of ferroelectric domains, and the imaging stage reads out the domain size. The piezoresponse can be measured during the DC pulse acquisition by superimposing the measuring AC sinusoid, or, alternately, immediately after each DC pulse (so the measuring sinusoidal waveform has no DC offset). These two situations are termed 'on-field' and 'off-field' (sometimes also termed 'in-field' or 'out-of-field'). The measured signal at each point can then be represented as a local on-field and/or off-field hysteresis loop, containing information on local ferroelectric properties and their evolution with bias. After the introduction by Hidaka *et al.*,[156] this technique was rapidly adopted by the community and extensively used to study polarization switching and spatial variations of ferroelectric behavior in PZT,[59, 133, 134, 157, 158] $BaTiO_3$,[159, 160] $Pb_{0.76}Ca_{0.24}TiO_3$,[161] and more.

This technique was taken a step further by the introduction of switching spectroscopy PFM (SS-PFM). In SS-PFM, the hysteresis loops are measured over a dense spatial grid of points, resulting in a 3D data set.[98, 162] Statistical or model-based analysis of the hysteresis loop then allows quantification of the relevant aspects of ferroelectric behavior, including the coercive and nucleation bias, remnant response, etc., and visualize them as 2D maps. This approach was used to visualize the nucleation centers in ferroelectric materials,[163] explore the interplay between the ferroelastic walls and switching further enabling observations and control of non-180° switching,[164, 165] observe collective dynamics during switching in large ferroelectric capacitors,[166, 167] determine the role of an isolated domain wall[168] and isolated single defect[169] on the switching process, and map the variation of switching behavior across a model grain boundary in a bi-crystal.[170] Several examples of where SS-PFM has been used are shown in **Figure 5**, along with the comparison from simulations provided by phase-field modeling. [171-175]

### II.c. Complex excitations and spectroscopies

The broad adoption of PFM by the ferroelectrics community for imaging, spectroscopy, and domain writing has further stimulated a question of whether more subtle aspects of ferroelectric behavior can be probed locally (or even have local analogs), including ferroelectric non-linearities[2, 20, 176, 177] and Preisach densities.[19, 20, 178-181] At the same time, a critical number of authors reported that despite the fact that generally PFM measurements yield numbers close to that expected for $d_{33}$ (or $d_{eff}$) of specific materials and the fact that PFM contrast is expected to be quantitative based on contact mechanics (or even simple dimensionality theory considerations), there is a strong evidence of topography related cross-talk and irreproducibility in quantitative measurements.

The latter behaviors were understood from the fact that measured signal in PFM is a convolution of electromechanical response of the surface and the transfer function of the cantilever. These frequency behaviors were explored by a number of authors.[95, 182-184] At low frequencies the frequency dispersion of the transfer function is relatively small and is primarily associated with the non-idealities of the cantilever design. Practically, variations of 10-20% (or more) are common, allowing establishing piezoelectric properties within this order of magnitude. Note that this issue was also explored in the context of non-contact techniques such as magnetic dissipation force microscopy,[185, 186] where small dispersion of cantilever transfer function precluded (at the time) quantitative studies.

The quantification of PFM signal becomes significantly more complex close to the resonances, where minute changes of contact conditions result in shifts of resonant frequency and correspondingly strong changes in the amplification factor. At the same time, these conditions offer the natural sweet spot for ferroelectric imaging. The classical SPM strategies based on phase feedback frequency tracking via phase locked loop[187] are not applicable in PFM, since the phase of the response depends on the local domain orientation. This also follows from the fundamental physics of SPM imaging, where the local driving force is unknown and hence the system is physically undetermined (two observables in single frequency detection vs. three unknowns even for simplest harmonic oscillator model of cantilever dynamics).

To address these limitations, several solutions were developed. The band excitation method (BE)[188-191] utilizes the excitation of the probe by the wave packet comprising multiple frequencies selected to encompass the resonance. The detected cantilever deflection (response) is then Fourier transformed, and subsequent fitting by a simple harmonic oscillator model yields the cantilever resonance frequency, quality factor $Q$, response phase and response magnitude. An alternative approach is based on amplitude-based resonance frequency tracking, embodied in the dual AC resonance tracking method (DART).[192] In this, the difference between the amplitudes to the left and right of the resonance frequency is used as a feedback signal allowing maintaining of the system at the resonance. Furthermore, mathematical analysis of the data under some general assumptions of the peak shape allows extraction of Q-factor. Finally, the fast lock-in sweep approach by Hurley[193, 194] and rapid imaging by Huey[183, 184] at sequential frequencies should be mentioned in this context. In all these cases, the electromechanical response and local resonance frequency can be determined simultaneously. This allows both cross-talk free quantitative electromechanical characterization, and also provides insight into the local mechanical properties of the surface similar to atomic force acoustic microscopy.[195-197] Notably, both BE and DART signals can be used to substitute the sinusoidal excitation signal in more complex PFM spectroscopies, giving rise to a set of advanced quantitative spectroscopies.[198]

Finally, the need to understand more subtle aspects of ferroelectric functionalities have given rise to a set of multidimensional spectroscopies exploring dependence of local response on driving amplitude to map ferroelectric non-linearities, time, time and bias, and first order reversal curve measurements. Many of these methods are summarized in Ref [198]. Note that these techniques necessitate the use of BE or DART to use resonance amplification and hence achieve measurable signal levels.

Overall the key drawback to most of the spectroscopic approaches is the acquisition time, which necessitates a trade-off in the resolution in either the spectral or spatial domain. Very recently, a subset of the current authors has employed full information acquisition (i.e., streaming and capture of the photodetector) through a "general-mode" technique wherein the acquisition rates are increased by orders of magnitude whilst maintaining good signal-to-noise ratios.[199, 200] The extension of the general mode technique to spectroscopies[201] should offer time and spatial resolution near the fundamental instrumental limits.

## III. FERROELECTRICITY IN NON-FERROELECTRIC MATERIALS

The emergence of PFM that allowed mapping of ferroelectric domain structures and studies of polarization switching in ferroelectric materials has resulted in significant progress in multiple areas of ferroelectric thin film research, as illustrated in Figure 1. In fact, most groups active in the field of ferroelectric research adopted PFM as a primary imaging and characterization tool, resulting in the explosive growth of the applications of this technique. Given the complexity of alternative methods for probing ferroelectricity on the nanoscale, PFM observations of the remnant states ("domain writing") and hysteresis loop measurements have often become *de facto* proof of ferroelectric behavior on the nanoscale. Note that additional evidence towards the existence of polar distortion can be obtained from optical second harmonic generation,[202] Raman,[203] or direct atomic-scale observation of polar distortions.[204-209] These techniques are generally more involved and also generally do not provide information on the presence of two or more (meta) stable polarization states. Further, the definition of ferroelectricity is one in which the polarization is re-orientable with applied fields,[210] and therefore the purely imaging modes are insufficient. While in certain cases the presence of topological defects such as vortices and domain walls in static images provides strong evidence towards the ferroelectric nature of a material, these studies are still not preponderant in the field. Furthermore, PFM imaging and spectroscopy can be performed on a broad range of non-crystalline and beam sensitive materials that are inaccessible for electron microscopy.

Motivated by the successes of PFM as applied to classical ferroelectric materials and also classical piezoelectric biological systems, a number of groups explored the piezoelectric and ferroelectric phenomena in a broad variety of systems ranging from tobacco mosaic virus,[148] yeast and red blood cells, to pig and aortic walls.[211, 212] In parallel, many groups have explored ferroelectricity in non-ferroelectric materials in which ferroelectricity can in principle be induced by nanoscale confinement or strain, or conversely explored the limits of ferroelectric phase stability under nanoscale confinement. The observations of ferroelectric-like behavior in the nanoscale, for systems which typically do not show such properties at the macroscale, are summarized in **Table I** below. Some examples from the literature of such reports are shown in **Figure 6**. Many instances of biological ferroelectricity were also reported, some of which are shown in **Figure 7**.

In most cases summarized in Table I, the PFM studies were used as evidence towards the ferroelectric nature of the material, and material specific theories were developed to explain the origins of observed behavior. In other cases, alternative mechanisms were considered, such as (but not limited to) examples in the work of Gautier *et al.*[213], who measured electromechanical response, switching loops and domain retention in amorphous $LaAlO_3$ thin films, Kholkin *et al.*[214, 215] (piezoelectric behavior in $SrTiO_3$ ceramics), or Sharma *et al.*[216] (switching behavior in $LaAlO_3$). However, the birds-eye view overview of the published results suggests the remarkable commonality between the observed phenomena. Indeed, only in a few cases were the observed phenomena roughly identical to classical ferroelectric behaviors, including observations of pre-existing domain contrast (not related with possible topographic cross-talk), binary polarization states, domain evolution via domain wall motion, and well-defined hysteresis loops. In many cases, non-classical ferroelectric behavior was reported, including a continuous spectrum of polarization values and relaxation via uniform contrast change, strongly reminiscent of relaxor ferroelectrics.[217-222] The hysteresis loops also rarely displayed saturation, unlike in a traditional ferroelectric. Remarkably, no clear delineation between these cases were found – classical ferroelectrics can demonstrate these abnormal contrasts, whereas many non-ferroelectric materials show remnant states and hysteresis loops.

In addition, several groups have recently reported the observation of pressure-induced switching in the ultra-thin ferroelectric thin films (see also Figure 6). In most cases the authors suggest that the reason for the observed switching is due to large strain gradients coupling with the polarization (the flexoelectric effect) leading to polarization reversal with pressure applied by the AFM tip.[223] However, other competing explanations, such as segregation of surface charges under pressure, were generally not considered in these works, and moreover, the difficulty in determining the flexoelectric coefficients by experiment are relatively well-known. As an example, the control experiment for separating this effect from others, as demonstrated beautifully by Damjanovic for single crystals,[224] reveals that chemical inhomogeneity within samples are sufficient to cause the accumulated charge that is typically used to determine the flexoelectric coefficients, resulting in errors in calculated coefficients orders of magnitude higher than their actual values.[225]

The above table and observations suggest that separation between materials that are not ferroelectric, from those that are non-traditional ferroelectrics, is not possible from standard PFM measurements. This has been recognized in the literature, and several possible solutions have been proposed by various groups. These are discussed further in section VI; below in section IV we first review some other related observations on oxide surfaces.

## IV. RELATED OBSERVATIONS

Like many other techniques, PFM provides only one aspect of information on ferroelectric behavior. After the initial discussions in the late 90s, it was generally accepted that the PFM signal is originating from converse piezoelectric effect, and subsequent technique development and interpretation of observed data was driven by this preponderant viewpoint.

However, this interpretation ignores the broad panoply of chemical, electrochemical, and physical phenomena emergent on ferroelectric (and non-ferroelectric surfaces). Early work on SPM of ferroelectrics had already contained sufficient indicators that in some cases, the origin of the observed contrast could be more complicated, as evidenced by the early work of Franke[43, 44] and Khim[45] postulating a strong electrostatic contribution to the contact-mode signal, work by Kalinin and Bonnell on Kelvin Probe Force Microscopy of ferroelectric surfaces during phase transitions and domain wall motion,[60, 61, 74-76] a set of work by a group at Argonne on chemical screening,[78, 79, 226-229] and certain PFM observations such as back-switching and formation of bubble domains.

The second element that is key to the interpretation of PFM data is the alternative origins of the electromechanical coupling that can contribute to the measured signal. These include electrostatic forces, non-piezoelectric electromechanical phenomena including surface piezoelectricity and flexoelectricity,[148, 149, 230] and surface- and bulk electrochemical phenomena. The electrostatic and electrochemical phenomena will be discussed in detail in Section V. We further defer the discussion of the role of flexoelectric coupling in PFM to a separate publication. We note that the flexoelectric effect describes the instant response of the polarization to the strain gradient in the system. It can give rise to additional electromechanical response (via coupling through electrostriction); however, its effect on switching, formation of long-lived states, etc. can be understood only by considering the interplay between flexoelectricity and screening phenomena.

Below, we briefly discuss some of these observations, since they provide the broader context of phenomena on ferroelectric surfaces that may provide contribution to PFM contrast via alternative electromechanical strain mechanisms or charging/electrostatic forces. We consider surface ionic screening of ferroelectrics, electrochemical strain microscopy, as well as bias-induced phenomena in the context of electrochemical writing of oxide surfaces.

### IV. A. SURFACE IONIC SCREENING

The intrinsic aspect of ferroelectric surface and interface behavior is polarization screening. By definition, polarization discontinuities at surfaces and interfaces lead to the formation of polarization bound charge, $\sigma = \boldsymbol{P} \cdot \vec{\boldsymbol{n}}$, where $\boldsymbol{P}$ is polarization discontinuity and $\vec{\boldsymbol{n}}$ is the vector normal to the interface. In the absence of screening, the bound charge creates an electric field across the material which suppresses the polarization. In classical ferroelectrics in which polarization is the primary order parameter, this decrease is of the order of ~10,000K, meaning that the ferroelectric phase is absolutely unstable. Correspondingly, the stability of the ferroelectric phase necessitates efficient screening.

While the necessity for efficient screening[231] was realized since the early days of ferroelectricity, the preponderant screening models evolved with time. The early literature in the field implicitly or explicitly assumed that screening processes happen on the surfaces and is irrelevant for bulk behavior, leaving the issue of an exact mechanism open. The work by Vul,[232]

Guro,[233] and Fridkin[234] explored the coupling between the polarization and intrinsic carrier dynamics in the material, giving rise to the ferroelectric-semiconductor concept. While a broad range of polarization-dependent photovoltaic and photochemical phenomena suggest that ferroelectricity indeed affects the near-surface band structure, simple examination of the corresponding length and energy scales suggests that intrinsic screening in a high-band gap material with low carrier concentration is relatively inefficient. Correspondingly, in the beginning of XXI century the main model for ferroelectric screening has become external screening by surface ionic charges, as explored by SPM,[74, 76] X-ray[78, 228] and density functional theory.[235] The direct visualization of the mobile surface charges was performed using time-resolved Kelvin Probe Microscopy.[236, 237]

The necessary consequence of the ionic screening model is that polarization switching necessitates the redistribution of the screening charges, and overall polarization dynamics is now controlled both by ferroelectric and ionic subsystems. This behavior resembles the electrochemical electrode polarization process in solutions.[238, 239] The second consequence is that polarization switching becomes a non-local process, necessitating lateral ionic transport between the regions of cathodic and anodic process to satisfy both charge- and mass balance. For example, in classical electrochemical systems with low ionic resistance it was shown that the macroscopic electrode separated by millimeters from the tip surface contact area controls the electrochemical reduction potential.[240] In the systems with low conductivity, the localization of the counter-reaction zone is determined by the interplay between the voltage drop and localization of opportunistic reaction sites,[241] for example water electrochemistry. For polarization, the underpinning phenomena are similar, except that regions with switched polarization play the role of a virtual electrode. The accumulation, injection, and transport of ionic charge now becomes an inherent part of the switching process, as illustrated by phenomena such as the formation of bubble domains during back switching,[80, 242] domain shape stability loss,[85] and chaos[86] during ferroelectric switching.

The important difference between classical electrochemistry and electrochemistry of a ferroelectric surface is the type of control. Classical electrochemical processes are voltage controlled, making them theoretically amenable for thermodynamic and rate theory studies and opening a ready pathway for experimental exploration in liquid- or solid state electrochemical cells with conductive electrodes.[243, 244] However, polarization switching is generally a galvanically controlled process, where a fixed amount of charge is transferred. The electrochemical potentials developing in the systems are controlled by the total driving force (the bulk polarization energy), availability of screening species and bias dependence of the screening processes. In the presence of efficient screening with metallic electrodes, the developed potentials are very small (and generally determined by the structure of dead layers). In the liquid environment or a "dirty" atmospheric environment with high ionic conductivity the developed potentials will be sufficiently large to ensure rapid electrochemical screening, with migration or diffusion to the surface or ejection of uncompensated charges as limiting step. Finally, in the clean environment where the amount of screening species is very limited, the galvanostatic control leads to the emergence of high potential enabling high-energy screening processes via electron and vacancy emissions. The evidence for the high energy of these processes is e.g. the X-ray emission from ferroelectric surfaces, which has been known since the late 80s.[245, 246]

While a full description of associated processes and in-depth of analysis of the dynamics still has not been achieved, the basic nature of the coupling between ferroelectric and electrochemical processes on surfaces is by now well established, and should be taken into account during interpretation of data.

### IV. B. ELECTROCHEMICAL STRAIN MICROSCOPY (ESM)

Electrochemical phenomena can be a strong contribution to the electromechanical coupling. Recently, it was proposed that application of electric bias to the ionically conductive material will result in ionic motion, and, via Vegard strain, will give rise to the surface displacement.[247] This concept was utilized in Electrochemical Strain Microscopy (ESM),[248-251] allowing visualization of phenomena such as electrochemical activity on cathode[252] and anode[253] materials, oxides[254-256] and semiconductors. Similar to PFM, ESM can be extended to a variety of time- and bias- spectroscopic modes.[248, 257] The original description of ESM mechanism considered purely diffusional dynamics with fast voltage controlled generation-recombination of ionic species on the surface and near surface regions.[258] Subsequently, the model was extended to include both migration transport[259-261] and electrostrictive effects.[118, 262, 263] Finally, it was shown that ESM mechanism can be present even in the absence of a surface reaction. In this case, the redistribution of ionic species within the material creates an electrochemical dipole, the coupling of which to electrostriction gives rise to electromechanical coupling.

Here, we note that the significance of ESM for PFM is that (a) electromechanical response can be observed on the non-ferroelectric materials and (b) electrochemical polarization in the solid electrolytes can give rise to long-lived polar states, with the characteristic life time determined by the ionic transport. In fact, from the point of view of external electromechanical measurements of a ferroelectric sample and an ionic conductor between blocking electrodes, these measurements are fundamentally indistinguishable. In the first case, the dipoles are confined to unit cells, whereas in the second case they originate from ionic motion between several unit cells.

### IV. C. TIP INDUCED ELECTROCHEMICAL WRITING

Finally, this discussion will be incomplete without considering the spectrum of possible bias-induced phenomena on surfaces beyond polarization switching. Since early days of scanning probe microscopy, it has been understood that application of relatively modest (several volts) potentials to the tip results in extremely high electric fields that can result in highly unusual electrochemical phenomena. Due to the (often) highly localized nature of the reaction process, it can be used as a basis for tip induced nanofabrication.[264] Well documented examples include nanooxidation of metals and semiconductors.[264-271] Remarkably, even such non-classical processes as formation of silicon carbide from silicon oxide and ethanol under 20V bias were documented.[272-276] Note that unlike classical electrochemistry in planar electrode cells (or devices), the electric field falls away from the tip surface junction, precluding classical electrical breakdown and enabling unusual reactions to occur at high biases.

The electrochemical reactivity of the oxide surfaces is actively explored in the context of solid state fuel cell applications.[243, 277] However, the macroscopic studies have necessarily been limited to the high-temperature conditions where ionic conductivity of the bulk is sufficiently high. At low T, the IR drop in the bulk dominates system response and electrochemical studies become impossible. In comparison, on the nanoscale the electrochemical and ionic processes are enabled at much lower temperatures, as analyzed in Ref.[278, 279] Notably, many phenomena such as unusual behavior of $LaAlO_3$-$SrTiO_3$ (LAO-STO) system[280] and domain wall conductance in ferroelectrics were originally interpreted in the context of purely physical models (polar catastrophe for LAO STO[280, 281] and band-gap lowering for domain walls[11]) and were further shown to have strong chemical components including vacancy dynamics and surface water electrochemistry (see refs [282-285] for LAO-STO and refs[286-288] for domain walls).

The natural question arises as to what conditions are required to induce electrochemical processes on oxide surfaces. The field strength alone does not provide the required information, since even application of 1 mV to nanometer size tips can generate fields of the order of 10 MV/m, well above the dielectric strength of many materials. Rather, the acting potential difference in the tip-surface gap should exceed the thermodynamic stability limit of material. The potential in the gap is determined by the (capacitive and resistive) voltage divider effect between the bulk and the material. Correspondingly, an approach to explore experimentally these phenomena is to study materials with high bulk conductivity, where the potential drop is localized at the junction. Recently, Tselev *et al*. performed numerical modeling of the electrochemical strain microscopy (ESM) signal generated from solid electrolytes, established the relevant time scales and frequency dispersion of the response.[289]

Experimentally, ESM studies by Kumar *et al*. have demonstrated that in materials such as $La_{1-x}Sr_xCoO_3$ the electromechanical hysteresis loops open at voltages of ~2.5 V without irreversible surface deformation.[254, 256, 290] At the same time, irreversible surface deformation and amorphization starts at ~10 V, establishing the lower boundaries for the reversible and irreversible processes respectively. Further insight into this behavior was obtained by Vasudevan *et al*.,[291] who observed formation of single oxygen vacancies on manganite surfaces with atomic resolution via first order reversal curve spectroscopy and atomically resolved scanning tunneling microscopy.[292] These studies suggest that ~2.5 V is the lower bias required for formation of oxygen vacancies on oxides. While the universality needs to be established, in both cases the potential is controlled by free energy of oxygen reduction/evolution reactions.

# V. MECHANISMS AND BIAS-INDUCED PHENOMENA IN AFM

Application of bias to the tip can introduce a broad range of physical, surface and bulk electrochemical phenomena, ranging in the degree of irreversibility, charge and matter transfer, and changes in surface and bulk composition. Here, we suggest the classification of these phenomena based on the nature of the material's changes. Given that the PFM/ESM measurements typically involve two steps – formation of the remnant electromechanically active states and electromechanical response of surface *per se*, we tailor the classification accordingly. Below, we provide a general overview of possible surface changes, and briefly discuss relevant physical mechanism including characteristic response magnitudes and the frequency dispersion of response.

## V.A. GENERAL CLASSIFICATION OF BIAS-INDUCED PHENOMENA

The phenomenological schematics of possible tip effects is illustrated in **Figure** 8. Here, the left-hand side summarizes purely physical mechanisms, associated with changes in materials structure on the unit cell level. These include the ferroelectricity (reorientation of dipoles within unit cell)[36, 37] and thermal expansion due to the tip-surface current.[293] On moving to the right, the degree of the changes in the system increase, with the ionic and electronic species redistribution on the length scale of multiple unit cells. These include the formation of electrochemical dipoles (electrochemical polarization), charge injection, and field-effect charging. Here, charge injection refers to direct injection of charged species from the tip to the material, with the sign of injected species being the same as the probe bias. The field effect generally refers to the accumulation of the charged species of the opposite sign to screen the tip bias, under the conditions where electronic and ionic transport between the tip and the probe is inhibited. We can classify the various mechanisms according to three changes– namely, change in concentration ($\Delta n$), change in charge ($\Delta q$) and change in dipole moment ($\Delta p$), as indicated in the figure. For example, the Vegard strain involves change in concentration, but no change in dipole moment or charge (as electroneutrality is maintained), whereas polarization switching in ferroelectrics does not involve change in concentration of any species, but does involve change in dipole moment.

Note that in classical semiconducting systems the phenomenology of charge injection and field effect is extremely well developed.[294] However, this description is typically limited to the electronic carriers in semiconductors, assuming that ionic motion does not occur and generally chemical degrees of freedom are frozen. Similarly, classical simple electrochemical descriptions often postulate local electroneutrality, and consider spatial charge separation only on relatively advanced levels of theory (often lacking simple analytical descriptions). In comparison, both conditions can be violated in the SPM experiment due to high field concentration and small length scales of the tip, and presence of opportunistic mobile ionic species on surfaces in ambient environment. Hence, in discussing the charge injection and field effect, we note that both can (a) include electronic and ionic species (or both), and (b) include surface and bulk of the material. Correspondingly, the description of these phenomena should ascertain the changes in both surface and bulk state. For example, in materials with a purely inert surface in the absence of mobile chemical species, possible processes include purely electronic charge injection/field effect in the surface states or bulk trap states; these types of phenomena are likely to be observed on layered ferroelectric materials[295-297] in vacuum when the surface state can be established by direct atomic observations. In comparison, ionically non-conductive but electronically conductive materials in ambient environments are likely to undergo surface electrochemical processes, potentially accompanied by partial charge screening by conductive electrons, as observed on conductive oxide surfaces.[298] However, for the vast majority of materials with poor, but non-zero electronic and ionic conductance, many of these processes can co-occur and in general will be difficult to differentiate. Furthermore, the time scales of the processes can be different – for example, rapid injections of the electrons in the surface water layer first leads to the formation of solvated electrons, and only later evokes electrochemical process and formation of hydrogen and surface hydroxyl species.

As a general guideline, measurements for different degrees of surface preparation or humidity (controlling the thickness of conductive water layers), exploration of model systems with varying degree of electronic conductivity (e.g. STO and Nb-STO), and the use of surface-sensitive chemical analysis tools (for irreversible changes) are possible ways to obtain insight into the observed phenomena. Note that while certainly complex, this situation is not unusual in material science and electrochemistry – it suffices to point to the ambiguities in the interpretation of impedance spectra of electrochemical systems.

A special limiting case of the reversible processes is the Vegard strain process, corresponding to the injection in the bulk of species generated at the tip-surface junction, corresponding to the bulk electrochemistry approximation. Finally, application of bias can lead to much stronger changes in material, including strong changes in chemical bonding or composition. These may include electroplating and electrodeposition, injection and ordering of vacancies, and surface and sub-surface damage. The localization of these damage sites can be both directly below the tip and at the interfaces (e.g. due to change in conductivity character). Generally, such changes are irreversible and are readily detected via comparison of topographic images before and after surface modification, or via chemical imaging tools such as microRaman, time of flight mass spectrometry[299], focused ion beam milling and scanning transmission electron microscopy,[300] or atomic probe tomography imaging of resultant changes.

### V.B. Physical processes detected and observed by PFM

We further proceed to briefly describe relevant physical mechanisms, including the magnitudes of surface deformation and time scales involved in formation of remnant states and frequency dependence of electromechanical response. When available, we further provide references to the detailed studies.

**Ferroelectric switching:** The most explored case of the tip-induced switching is the classical ferroelectric material. In this case, the emergence of remnant states correspond to formation and growth of classical ferroelectric domains, well explored both analytically[301-304] and via phase field modelling.[305, 306] The general features of this process involve nucleation above certain threshold voltage and subsequent growth with time and bias. The extant domain tends to remain constant (if strongly pinned) when the bias is switched off, or relax through lateral motion of domain walls. Due to the importance of this process to the field of ferroelectricity, multiple studies of this process as affected by localized and extended defects, screening, switching in non-tetragonal material and non-180° switching, etc. have been reported. Additionally, switching dynamics in the vicinity of domain walls,[307, 308] switching of more complex ferroelastic domain patterns,[309-311] switching near extended defects,[169, 170] and as a function of humidity[312-314] have all been explored. Here, we will not recap these studies but note that they indicate the formation of a well-defined domain about a certain threshold voltage, with domain growth through lateral domain wall motion. More complex phenomena are possible for multidomain materials and domain textures; however, these main features are still present.[310, 315-320]

The (fundamental) detection mechanism for these systems is the converse piezoelectric effect. Here, the strain (and hence surface displacement) is $U = QP^2$, where $Q$ is electrostrictive coefficient and $P$ is polarization. For a ferroelectric, $P = P_{ferro} + \varepsilon\varepsilon_0 E$, which yields piezocoefficient $d_{eff}$ for 1D case as $d_{eff} = \varepsilon\varepsilon_0 Q P_{ferro}$ where $\varepsilon$ is the dielectric constant and $\varepsilon_0$ is the vacuum permittivity. The important consequence is that the presence of only two response levels corresponding to $+P$ and $-P$ states are characteristic features of this mechanism. For a realistic SPM configuration, the magnitude of the response was extensively analyzed and coupled,[100-102] numerical,[108] and approximate[109-112] solutions are well known. For intrinsic switching, the response time is below nanoseconds, i.e. instant from the SPM perspective. The situation can be more complex for the cases involving domain nucleation and wall motion controlled phenomena; however, these can be readily resolved via observations of pulse-induced changes in domain structure or in lateral field experiments.[312, 319, 321-325]

Note that under realistic conditions the additional contribution to the signal originates from electrostatic forces. However, the magnitude of the latter depends on the screening conditions (i.e. is non-universal) and for strong ferroelectric materials these tend to be considerably weaker than the piezoelectric effect.[94]

**Vegard strain:** The second most well studied case is the Vegard strain mechanism, initially developed to quantify electrochemical strain microscopy (ESM).[191, 247, 248, 252, 326] Here, it should be noted that Vegard strains in most materials (10-100%) are significantly larger than piezoelectric and electrostrictive ones (0.01 -0.5%), necessitating consideration of this mechanisms. Here, the reaction in the tip-surface junction is assumed to result in the generation and recombination of ionic species that subsequently are transported in the solid. For purely diffusional transport, the analytical solution was developed for 1D[247] and rotationally invariant 3D cases[258] and approximate solutions were developed for the anisotropic case.[327] In this case, surface displacement is proportional to $\beta D_{ion}/\sqrt{\omega}$, where $D_{ion}$ is the diffusion constant of the mobile species, $\beta$ is the Vegard coefficient, and $\omega$ is frequency. The characteristic time scale of the diffusion process is $R^2/D_{ion}$, where $R$ is the characteristic system size. This suggests two important limits, namely a high frequency regime where diffusion length is smaller than the tip radius and process is essentially 1D, and a low frequency regime where the spatial extent of affected region diverges with time. ESM was extended to regimes with both diffusive and migration transport for open and blocking[259] electrodes, giving rise to more complex time dispersion of responses.[289, 328] The important aspect of pure Vegard mechanisms is that it does not predict the formation of remnant states with opposite electromechanical responses and hysteresis loops beyond kinetic ones. Combined with the explicit assumption of fast surface reactions, these mechanisms can be relegated to purely electrochemical systems under rapid electrochemical control, e.g. low frequency dilatometry[329, 330] or dynamic response sin system with highly-mobile ions and fast surface reactions.

**Polarization/electrochemical dipoles:** The third well studied case is the formation of electrochemical dipoles on the (ionically blocking) surfaces. While the mechanisms of this process have not been (to our knowledge) explored in depth for oxide surfaces, the basic physics of these phenomena has been well studied in the context of classical electrochemistry.[238, 331, 332] In this mechanism, the application of electric bias to the tip in contact with the surface polarizes the mobile ion sub-system, resulting in classical double layers. Low kinetic mobility results in high remanence of these states (somewhat similar to electrets), with characteristic relaxation times of the order of $R\lambda/D$, where $R$ is the characteristic system size, and $\lambda$ is the Debye length. The fast electromechanical response originates from the coupling between the electrochemical dipole and electrostriction coefficient, i.e. $d_{chem} = \varepsilon\varepsilon_0 Q P_{chem}$, with characteristic response times on the order of nanoseconds.

**Charge injection/field effect:** For the case of electronic and ionic charge injection the mechanisms become more involved, since they now involve the transport and relaxation of multiple species. While the SPM based theory is absent, more general considerations suggest that characteristic timescales will be comparable to the polarization case (from simple dimensionality theory). However, the presence of uncompensated charge will result in strong electrostatic interactions, as will be discussed below.

**Heating:** For the case of significant surface conductivity additional contribution to electromechanical coupling is possible from tip-surface current.[333] While the surface expansion is quadratic in voltage, the surface change and non-linearity leads to rectification and emergence of the first harmonic of response, phenomenologically similar to the piezoeffect. The basic physics for Joule expansion is similar to the Vegard case, since both transport of ionic species and heat follows a diffusion equation. However, unlike the Vegard case, the absolute value of current can be measured independently, obviating the need for ad-hoc assumptions on the generation process. The sample expansion from the Joule heating effect can be written as $U = \beta R I^2$ where $\beta$ is the Joule heat transduction coefficient I is the measured current and R is the resistance. By measuring the current at the same time as acquiring the hysteresis loop, the contribution to the sample expansion from the Joule heating effect can be estimated.[333] Practically, for materials with effective properties of oxides these contributions are typically negligible if the resistance is $>\sim 1G\Omega$.

Overall, the summary of possible mechanisms is given in **Table II**. Note that analytical descriptions are generally available only in cases when a single response mechanism is operational, including piezoelectric response and ferroelectric switching, Joule heating, electrochemical polarization and electrostrictive response. However, in most cases response are complex and determined by the interplay of physical and chemical processes, in which cases only leading order approximations can be established and generally coupled numerical modeling is required to establish precise characteristics of the process.

## VI. BUT IS IT PIEZO/FERROELECTRIC?

Given the multitude of mechanisms giving rise to an electromechanical response, numerous groups have sought to clarify the origins of the measured deflection signal to segregate the piezoelectric (and/or ferroelectric) properties from those of other origins. Below, we review some existing techniques, and then focus on a new technique termed contact Kelvin Probe Force Microscopy, that shows promise in distinguishing between ferroelectric and non-ferroelectric hysteresis. We finish by noting the strides made in quantifying the electrostatic and piezoelectric contributions, and briefly discuss possibilities of using acoustic detection in AFM studies.

### VI.A. MEANS TO DISTINGUISH PIEZOELECTRIC FROM NON-PIEZOELECTRIC RESPONSE

Several approaches have emerged to attempt to decouple ferro/piezoelectric from non-ferro/piezoelectric sources in the electromechanical response. As a guide for the researcher, we present in **Figure 9** a list of possible mechanisms that can contribute to electromechanical response and (sometimes) hysteresis, as well as their typical experimental features. For example, observation of a clockwise hysteresis loop is not a feature of a standard ferroelectric, and is strongly suggestive of charge injection or other phenomena. Alternately, formation of electrochemical dipoles is associated with a continuum of states in the contact KPFM response, a feature that is absent in standard ferroelectrics with two stable polarization states. Although the existence of non-ferroelectric signal contribution in SPM-type experiments is widely acknowledged, clear guidelines on how to avoid or distinguish them from ferroelectric responses has been lacking.

One example, as explored by Strelcov et al.[334], is the use of varying $V_{ac}$ amplitudes during the hysteresis loop acquisition. For a typical ferroelectric, the amplitude of the applied AC voltage should not correlate with the height of the acquired (off-field) hysteresis loop so long as $V_{ac} \ll V_c$ where $V_c$ is the coercive voltage (i.e., the voltage at which polarization reversal becomes inevitable in the quasi-static case). As $V_{ac}$ increases, the loop becomes progressively more narrow, until the loop collapses almost entirely once $V_{ac} \gg V_c$ with a corresponding collapse in the loop height around $V_c$. This is possible in true ferroelectrics because applying a $V_{ac}$ beyond the coercive field corresponds to switching during the measurement cycle (i.e. intrinsic switching is very fast compared to PFM time scales), so total hysteresis measured is effectively zero. Absence of such dependence and loop collapse is indicative of non-ferroelectric contributions to the observed electromechanical hysteresis. This technique is shown in **Figure 10 (a,b)**.[335]

Alternatively, one can also explore the frequency dependence of the signal, a technique explored recently by Kim et al.[336] Here, the authors exploit the fact that across a frequency range of a few hundred kHz, the (intrinsic) piezoelectric properties of ferroelectrics are not heavily dependent on the probing frequency, whereas they are likely heavily frequency dependent for ionic redistribution. The surface displacement induced by ionic motion can be modeled as

$$\frac{u_3}{V_{ac}} = \frac{2(1+\nu)\beta\sqrt{D}}{\eta} \frac{1}{\sqrt{\omega}} \qquad (2)$$

where $u_3$ is the surface displacement, $\nu$ is Poisson's ratio, $D$ is the diffusion coefficient of the ion, $\eta$ is the linear relationship between applied electric field and chemical potential, and $\omega$ is the frequency of $V_{ac}$, suggesting that the frequency dependent sweep can determine whether this process is contributing to the observed electromechanical response. The authors could find that there was no such relation for PZT in the range 10-100kHz, but that substantial dispersion was found for LICGC (a battery electrode material), providing a promising route towards appropriate signal classification (see **Figure 10(c,d)**). Note that this approach has to be modified for polycrystalline ferroelectrics and especially capacitor based devices, where frequency dispersion can emerge as a result of domain wall dynamics within probing volumes.[337-343] However, the latter is usually reasonably weak (~<10%/decade).

Recently, combined multimodal techniques have also been suggested, for example by Li et al.[118] who utilized higher harmonic excitations and dependence of hysteresis on voltage range and frequency of the applied DC waveform, to separate electromechanical response from spontaneous polarization from those stemming from induced dipoles. One can write the electromechanical response as proportional to $P_S^2 + 2\chi P_S E + \chi^2 E^2$. The argument is that the electrostrictive contribution to the traditional ferroelectric is small and hence the sample deformation remains mostly linear with AC fields (below about half the coercive field), as the spontaneous polarization is large. On the other hand, if the induced polarization is large and the spontaneous polarization is small or non-existent, then the second harmonic response should be large (a feature which is also seen in relaxor ferroelectrics). It was also found that there existed a substantial dependence of the switching on time scales (in the 1-0.1Hz range), as well as dependence of the coercive voltage on the voltage window used for soda-lime glass, both of which are characteristics of ionic motion and not observed in classical ferroelectrics (see **Figure 10(e-h)**).

Ultimately given the large number of possible mechanisms that can result in an observed electromechanical response, abundant caution is required during interpretation of results. The problem is also more challenging than simple shifting of time scales; for example, one prevailing method of differentiating the ferroelectric from non-ferroelectric sample is to observe the decay in the electromechanical response after applying a DC voltage. For typical ferroelectric materials (not relaxors), relaxation should be minimal with time.[335] However, charge injection can result in long-lived states with relaxation times that are extremely long (e.g. hours or even days, and even persisting at high temperature[344, 345]), and therefore, these also cannot be used as unambiguous proof of ferroelectric switching.

Fundamental progress requires quantification of the various contributions to surface displacement. Below, we discuss two major thrusts of research stemming from recent work on contact electrostatic measurements, as well as quantitative characterization of the surface displacements from both electrostatic and piezoelectric effects. We suggest these are methods through which the presence (or absence) of 'true' ferroelectricity can be discerned from those of 'strange' ferroelectrics.

**VI.B. CONTACT KELVIN PROBE FORCE MICROSCOPY**

Recently, we have shown that it is possible to directly and quantitatively measure electrostatic tip-sample interactions in the measurement compatible with piezoresponse force microscopy.[99, 335, 346] A straightforward way to separate these contributions is to consider potential-dependence of forces acting on the junction in contact, i.e. to consider the so-called on-field observables. Ironically, the original motivation to pursue off-field observables, such as remnant piezoresponse, was motivated the desire to minimize electrostatic contributions to the signal. As it turns out, in such measurements electrostatics plus charge injection from the tip alone can produce hysteresis loops that become indistinguishable from weak ferroelectricity with small spontaneous polarization. The origin of this phenomenon is deeply rooted in the propensity of almost any material toward defect chemistry and associated creation of deep and ambipolar electronic states that create potential imbalance in the junction, and the ensuing electrostatic force.

From the most basic considerations, the electrostatic force $F_{ES}$ is proportional to the square of applied voltage: $F_{ES} = \frac{1}{2}k^{-1}C' \cdot V^2$ and inversely proportional to the stiffness of the cantilever, $k$. In contact, the capacitance gradient $C' = dC/dz$ is poorly defined, compared to for example non-contact electrostatic force microscopy, and $C'$ has to be viewed as a proportionality coefficient between the force and voltage with local and global contributions from the tip apex, tip cone, and cantilever shank.[347, 348] Piezoelectric displacement is linearly proportional to applied voltage $F_{PE} = d_{eff}V$. The forces are probed with a sinusoidal voltage $V_{ac}$ superimposed on a dc bias $V_{dc}$,

$$V = V_{dc} + V_{ac} \cdot \sin(\omega t) \tag{3}$$

The PFM detection scheme measures the amplitude of the first harmonic of the total measured signal $D_{ac}$, which is proportional to the first bias-derivative of the total force. The derivative signal from the sum of electrostatic and piezoelectric contributions reads:

$$D_{ac} = \frac{d}{dV}F_{tot} = \frac{d}{dV}F_{ES} + \frac{d}{dV}F_{PE} \tag{4}$$

with

$$D_{ac} \approx C'V_{DC} + d_{eff} \tag{5}$$

We immediately observe that the electrostatic contribution to the measured displacement $D_{ac}$ depends linearly on the *DC* potential on the tip (which is a sum of applied potential and built-in surface potential, $V_{DC} = V_{tip} + V_{sp}$), whereas the piezoelectric contribution does not. The separation between ferroelectric and non-ferroelectric states in this case is efficiently carried out with contact electrostatic measurements discussed in Ref. [335].

Detection and subsequent decoupling of the DC-bias-dependence of the PFM signal can be carried out in a variety of approaches. The simplest qualitative method is just to compare on-field and off-field PFM hysteresis loops as shown in **Figure 11**. As a reminder, the on-field loops are obtained by applying $V_{dc}$ and $V_{ac}$ simultaneously and measuring the AC deflection $D_{ac}$. The DC voltage is varied in a series of voltage pulses with varying magnitude. The off-field hysteresis loop records $D_{ac}$ after the application of $V_{dc}$ to detect remnant changes induced by the DC voltage. The comparison of on-field and off-field PFM hysteresis loops for ferroelectric PZT and non-ferroelectric $HfO_2$ samples is shown in Fig. 11(a,b). Qualitatively, one observes that the off-field hysteresis loops of PZT has a significantly larger opening compared to $HfO_2$, but this really depends on the ferroelectric properties, i.e. the piezoelectric coefficient. The on-field hysteresis loops in Fig. 11b are likewise quite different, with the non-ferroelectric sample having a strongly sloped hysteresis. This indicates that electrostatic forces are the dominant signal origin which show an almost linear dependency on $V_{dc}$ and a small hysteretic part with unknown origin.

The more complex qualitative method to measure the DC-bias-dependence of the PFM signal was introduced through a technique called contact KPFM (cKPFM[346]). The disadvantage of just measuring the hysteresis loops in a spectroscopic manner as described above is that the fact that the remnant state which is altered by the DC voltage is only probed at $0V_{dc}$. However, in order to identify and differentiate some of the signal generating mechanism as listed in Fig. 8 and 9, it is of high interest to study the voltage-dependence of the bias-induced state. Therefore, cKPFM was introduced to probe the small amplitude DC voltage dependence after large amplitude DC voltage pulses. This can be done with a linear DC voltage sweep or spectroscopically as described in detail in ref[346]. Here, the small amplitude DC voltage needs to be small enough to not change the bias-induced state of the material and is often denoted by the parameter $V_{read}$ to differentiate it from the large amplitude DC voltage pulses $V_{write}$.

The DC waveform for a standard switching experiment is shown in **Figure 12**(a). We note that the 'read' step here is usually done during the $V_{dc}= 0V$ steps, i.e., the off-field loops, typically to minimize the electrostatic contribution. But to measure the signal as a function of this contribution, we proposed to vary the reading voltage within a small window. Therefore, the measurement now consists of application of $V_{write}$ pulses, followed by reading steps at different $V_{read}$ values. This is generally done in a sequential manner, so that the hysteresis loop is acquired at one particular $V_{read}$, then the loop is acquired again at the next $V_{read}$, and so on, and is indicated schematically in Fig. 12(b).

The cKPFM approach was applied to a variety of samples, with results shown in Figure 12(c,d). Figure 12c shows off-field hysteresis loops from three perovskite samples – ferroelectric PZT, $SrTiO_3$ grown on (110) $NdGaO_3$ (NGO) and an amorphous $HfO_2$ thin film. A cursory inspection reveals that all three have reasonably well-behaved hysteresis loops, with phase reversal and butterfly amplitude loop. In fact, similar hysteresis loops are typically what is reported in most publications. However, when the cKPFM signal is also explored, the picture changes drastically as demonstrated in Figure 12d. Each line corresponds to a PFM response as a function of $V_{read}$, recorded after larger DC voltage pulses color-coded with the side-bar.

Remarkably, although the remnant hysteresis in Figure 12c are similar for the three perovskites, the behaviors in Figure 12c and 12d are clearly differentiated: bias-dependence of $D_{ac}$ signal on PZT is non-linear with a clear saturation in the cKPFM curves. Positive and negative voltage pulses change the domain orientation underneath the AFM tip and the collection of cKPFM curves resemble the ferroelectric hysteresis. In contrast, the $HfO_2$ sample reveals smooth, linear dependence of the PFM signal on $V_{read}$, with no saturation. The case of STO/NGO (middle panel) is more interesting as it reveals the properties of both PZT and STO/NGO, likely signifying ferroelectric component in this sample as evidenced in earlier publications. Therefore, cKPFM may be used to determine the existence of two stable remnant states, as opposed to a continuum of states that results from surface charging effects, and thereby strongly support (or discount) the presence of ferroelectricity in thin films and nanostructures.

Here, we note that DC voltage pulses for the non-ferroelectrics case simply offset the lines along the *y*-axis. The DC voltage at which the mechanical displacement is nullified corresponds to the surface potential, i.e., one can determine $V_{sp}$ simply by inspecting the intersection of the lines with the *x*-axis in the case of the $HfO_2$ sample as is typically done for traditional non-contact KPFM. At the same time, the intersection with the *y*-axis at $V_{read} = 0V$ is also observed to be varying. Indeed, this is the reason that an off-field loop can be measured. The bias-induced change in $V_{SP}$ is due to charge injection from the AFM tip itself, and is a complex process of charge injection and charge relaxation. As an example, we plot the cKPFM response from a single location in an $HfO_2$ amorphous thin film sample in **Figure 13**. The trace along the $V_{read} = 0V$ direction, which corresponds to a standard PFM hysteresis loop acquisition, is plotted in Fig. 13c. At the same time the trace along the $V_{read}$ axis, which corresponds to surface potential is shown in Fig. 13b. Therefore, cKPFM provides both information from complementary techniques, namely, the surface potential and the field-induced electromechanical response. Further, the contact capacitance gradient *C'* can be plotted as a function of the surface potential, as in Fig. 13(d). Remarkably, C' and $V_{sp}$ can be changed by an applied electric field in the chosen case of non-ferroelectric HfO$_2$, with surface potential showing smooth hysteresis between -0.5 and 2.5 V (Figure 13b), with the slope (proportional to the capacitance gradient) peaking around contact potential of 0.5 V (Figure 13d). Interpretation of these values is notoriously difficult in dielectrics, and can be carried out with continuum modeling of charge traps that are charged and discharged by applied electric fields.

In the following we will discuss more about the electrostatic forces and surface potentials in contact mode and the consequences on PFM imaging. For this, we again turn to amorphous thin film HfO$_2$ which has well known charge trapping properties and is not ferroelectric. Charges were injected form the tip by applying positive and negative voltages to the scanning tip in the same way ferroelectric domains are poled. The presence of surface charges is measured by traditional non-contact KPFM using a closed loop feedback. The contact surface potential is measured simply as an intercept of the cKPFM curves with the *x*-axis as explained above. As seen in **Figure 14**a-c, the contact and non-contact values show qualitatively the same image but are slightly different quantitatively. This originates from the influence of the lift scan height on the measured values for KPFM and the fact that in contact mode the surface potential is actually a junction potential between tip and sample. Due to the biased-induced surface charges and the presence of electrostatic tip-sample interactions, corresponding PFM signal, hysteresis and all the other properties can be derived. We also note here for comparison, that the PFM image from the poled areas for the HfO$_2$ thin film are shown in Figure 14d,e look markedly similar to PFM images taken on a poled ferroelectric sample (in this case, a PZT thin film), as shown for comparison in Figure 14f,g. Therefore, such contrast cannot be purely ascribed to "ferroelectric" nature of samples.

In summary, contact electrostatic measurements simply reveal the dominant origin of the measured PFM signal and allows firmly establishing the presence of bistability in the polarization states. Note that this cannot be replicated in classical PFM measurement, as these can show hysteretic loops due to charge injection from the tip or the continuous formation of electrochemical dipoles (either in the tip-surface junction, or in the material itself). In principle, ferroelectric switching can also change electrostatics through the change of $V_{sp}$. And for weak ferroelectrics with small value of piezoresponse, hysteresis of electrostatic response may provide a useful measure of ferroelectric properties. However, caution should be used to interpret hysteresis loops of predominantly electrostatic origin, because in this case PFM does not detect piezoresponse, and the so the implied equivalence of PFM signal and piezoelectricity cannot be established within the noise level of the measurement. Ferroelectric properties in this case need to be confirmed from other measurements, such as second harmonic generation and susceptibility measurements.

**VI. C.** DIRECT QUANTITATIVE MEASUREMENT OF THE NON-PIEZOELECTRIC CONTRIBUTION TO THE PFM SIGNAL

An obviously important point is the sensitivity of PFM measurements and contact electrostatics to the specific properties of the AFM cantilever. Any discussion of the role of electrostatics in piezoelectric measurements begs the question as to whether this additional contribution to the measured displacement can be quantified and therefore accounted for. A subset of the authors has recently shown that it is not only possible to account for the properties of the cantilever, but it is necessary to do so if quantitative values of the electromechanical response are of interest. This is especially true for resonance-enhanced measurements, where the common force-distance based calibration of the detector sensitivity in the atomic force microscope yields inaccurate results. The systematic procedure detailed in reference[349] takes into account the detailed shape of the eigenmode that is excited on resonance, the cantilever geometry, its spring constant and several other accountable properties. The cantilever dynamics can be largely accounted for in the so-called shape factor $\xi$, which allows to correct the experimental data to make the cantilever independent, and subsequently allowing quantitative comparison of the displacements measured with different cantilevers.

The developed calibration procedure was verified using a periodically poled lithium niobate (PPLN) crystal. In PFM, the dynamic cantilever displacement $D_{ac}$ is caused by a local sample volume expansion resulting from the inverse piezoelectric effect and is proportional to the piezoelectric constant $d_{eff}$. In our case, $D_{ac}$ is synonymous with the so-called PFM signal, which is typically a sum of the displacement from electrostatics, $D_{ES}$, and the displacement from the piezoelectric effect, $D_{PE}$ as described in Equation 1. Since $d_{eff}$ is a material property, the cantilever displacement for $D_{PE}$ should be independent of the cantilever properties.

Measurements on the PPLN crystal were performed with two different cantilevers, A and B, of different stiffness. **Figure 15a** shows the on-field PFM loops for the two cantilevers measured on a 128x64 grid over a 50 μm x 50 μm area and averaged over oppositely oriented domains before any data quantification. The plots are straight lines without any hysteresis. As expected, the softer cantilever shows a larger slope corresponding to a higher sensitivity to the electrostatic forces. This is due to the influence of the contact stiffness $k^*$ to the measured electrostatic contribution (see Equation 1). Based on the cantilever geometry, free as well as contact resonance frequencies, $Q$ factors, contact stiffness $k^*$ and the corresponding shape factors $\lambda$, surface displacements in pm were calculated for each point of the grid. As seen in the histogram in Figure 15b, the shape-factor calibration yields consistent results for both cantilevers (bottom panel), and this is not the case when using conventional calibration using only the static sensitivity $S$ (top panel).

After quantification of the PFM loops using the shape factor $\xi$ and the measurement of $V_{SP}$, the contributions from electrostatics, $D_{ES}$, and piezoelectricity, $D_{PE}$, can be separated and plotted, as displayed in Figure 15c,d. Figure 15c shows the electrostatic contribution for both cantilevers. As can be seen, $D_{ES}$ can be between 0.5 and 2 pm depending on the surface potential. Further, we compare $D_{PE}$ with the piezoelectric constant value $d_{PPLN} = 7.5$ pm/V specified by the PPLN crystal provider (Fig. 15d). It is seen that both the cantilevers yield values approaching the manufacturer-specified macroscopic one for domains of both orientations, and the results are independent of the surface potential. This shows the potential for PFM to become a quantitative characterization technique when all signal-contributing mechanism and the cantilever dynamics are considered. It also demonstrates a problem arising for weak ferroelectric materials with low piezoelectric constants. For $V_{ac} = 2$ $V_{pp}$, the electrostatic contribution was evaluated to be between 0.5 and 2 pm. If a piezoelectric constant would be one order of magnitude lower, for example, the piezoelectric contribution would be in the same range as the electrostatic contribution and, hence, would be difficult to detect.

The quantification of the electromechanical displacement opens pathways for identification of possible behaviors based on the strength of PFM signal. For example, for most materials the effective piezoelectric coefficient values are well known, and can be corrected for size effects by introducing (estimated) correlation lengths. Importantly, these values are independent on contact radius (i.e not sensitive to tip geometry). For electrostatic forces the situation is more complex, but for strong contact (range of significant electrostatic forces is smaller than contact radius) the response can also be estimated given the observation of maximal fields in the junction, suggesting tremendous opportunities for further studies.

## VI. D ACOUSTIC DETECTION

The insight into the nature of the tip-bias induced phenomena can be obtained from the introduction of a complementary measurable beyond the magnitude of the electromechanical response. In particular, a convenient observable in contact dynamic SPMs is the contact resonance frequency of the cantilever, directly linked to the contact stiffness of the tip-surface junction. The associated theory was developed in the context of atomic force acoustic microscopy (AFAM)[196, 350-353] and later adapted to PFM.[333, 349, 354] For ferroelectric materials, classical 180° domain switching does not change the elastic properties of the system. However, non-180° domain switching, tip-induced phase transitions, or electrochemical processes all are expected to shift resonance frequency due to a change in mechanical properties. Hence, observation of dynamic resonant frequency changes during PFM spectroscopy can provide additional insight into the mechanisms of the bias-induced processes.

In single frequency PFM the resonance frequency cannot be determined independently due to fundamental under determination of the system. The variation of resonance frequency due to surface topography leads to the cross-talk between topography and PFM signals. However, emergence of amplitude based feedback and multiple frequency methods such as band excitation[188, 190] has allowed to separate the electromechanical signal and resonant frequency both in imaging and spectroscopic PFM, as originally demonstrated by Maksymovych.[355] In the case of electrochemical reactions, Papandrew et al.[356] noted the strong change in resonant frequency during bias application in $CsHSO_4$ (CHS), coincident with reduction currents. CHS is a proton conductor at high temperature (above 140°C), but the changes in elastic modulus with voltage were absent in the room-temperature phase, or when non-catalytic probes were utilized, suggesting acoustic detection can be used as an effective probe of electrochemical reactions.

For ferroelectric systems with purely ferroelectric domain walls, application of DC bias and acquisition of hysteresis loops generally will not result in changes in elastic modulus. However, in ferroelastic systems, polarization switching involves the motion of non-180° domain walls, and therefore, can result in changes in the elastic modulus. Moreover, these effects are substantial in the case of field-induced phase transitions, when the elastic compliance diverges as the structural phase transition is approached, and have been observed experimentally.[357-360] For example, Li et al.[359] showed that for strained $BiFeO_3$ thin films, a transition from the rhombohedral to tetragonal phase could be induced at high fields, which could further be correlated to increased dissipation ($\approx 1/Q$), presumably from the motion of the interfaces in the probed volume of the tip. These results are outlined in **Figure 16.** Here we note that both intrinsic (structural changes) and extrinsic (e.g., domain wall or phase boundary motion) can have a substantial impact on the elastic properties, as motion of highly strained domain walls will manifest as large softening[361], and can be an indicator for phase transitions. Similar increase in dissipation was observed for mixed-phase $BiFeO_3$,[362] and which also correlated to the onset of a larger second harmonic response, reasoned to be due to the reversible motion of the phase boundary after a de-pinning transition. Naturally, the limitations for acoustic detection are that it is highly susceptible to small variations in contact area between tip and sample, and therefore, accurate determination of elastic modulus in the contact mode geometry requires full knowledge of the tip-surface contact. However, there are methods are being developed that may overcome this challenge.[363] Moreover, not all ferroelectric systems will exhibit the field-induced phase transitions that allow for direct observation of the softening, although an alternative route is the ferroelectric phase transition itself (presuming it is at an accessible temperature for the AFM).

# VII. COUPLING BETWEEN FERROELECTRIC AND IONIC/ELECTRONIC PHENOMENA

Finally, of special interest are the novel phenomena enabled by coupling between the electrochemistry and ferroelectricity. To provide some historical perspective, we note that initially interest to ferroelectric materials was driven by pure electromechanical, electrooptical, or ferroelectric functionalities. Electromechanical coupling enables applications in sensors and actuators, starting from celebrated SONARS and to the modern MEMS systems and smart dust. Electrooptical properties are explored in a broad range of optical devices, period doubling systems, modulators, etc. Finally, the presence of two or more equivalent polar states underpins applications in ferroelectric memories.[3]

The further development of materials science and condensed matter physics brought forward the interest in coupled functionalities. One example of such phenomena is the bulk magnetoelectric coupling.[364, 365] Another example is the coupling between ferroelectricity and electronic transport in the tunneling barriers[366-371] and field effect geometries.[372-374] Note that these functionalities are now spatially distributed and emerge as result of surface and interface effects coupled to the bulk ferroelectric phase behavior. Similarly, magnetoelectric coupling can emerge at the interfaces.[375-377]

Finally, in the last several years progressively more attention has been paid to the coupling between physical and electrochemical phenomena in oxides, especially pronounced in nanoscale systems[278, 279, 378-381] Similarly to the magnetoelectricity and electronic transport, the coupling between ferroelectric functionality and (electro)chemistry can be bulk and interface driven. In the bulk, Morozovska[382] analyzed the case of Vegard strain changing the sign of the first term in the Landau expansion, resulting in a ferroelectric phase transition. Recently, the attempt to control such states via He atom implantation was made by Ward *et al.*, who demonstrated that such strain control modulates the metal-insulator transition in $La_{0.7}Sr_{0.3}MnO_3$[383], and can also tune the structure of $SrRuO_3$ films between orthorhombic and tetragonal phases.[384]

Finally, an extremely broad range of emergent behaviors can be anticipated due to coupling between surface electrochemistry and bulk ferroelectricity. Indeed, ferroelectric phase stability in proper ferroelectric (i.e. polarization is the primary order parameter) necessitates screening of polarization charges. While this constraint was understood from the early days of ferroelectricity, it was believed that domain formation and some undefined screening mechanisms will essentially cancel depolarization fields in the bulk, ensuring the stability of ferroelectric phase. In the latter 1980s, a number of groups have explored the coupling between the ferroelectricity and semiconductor subsystem, giving rise to the concept of ferroelectric semiconductor.[234, 385-387] In parallel, the properties of spatially separated ferroelectric-semiconductor systems were explored as driven by FeFET and FeRAM applications.[388-393] Much of these results are summarized in an excellent recent review by Hong et al.[393] However, recent studies have demonstrated that in ambient conditions the screening is ionic in nature, and hence polarization dynamics at open surfaces should be considered as coupled surface electrochemical-bulk polarization switching process, as discussed in Section IV. Surprisingly the theory for such coupling was developed by Stephenson and Highland only recently,[229] and remained largely unused. Of course, control over these states requires a delineating the mechanisms that lead to observed electromechanical response in PFM, and given that traditional PFM cannot distinguish between the plethora of possible electromechanical response mechanisms, which have been described above.

## VIII. SUMMARY

The exquisite sensitivity of scanning probe microscopy to minute surface displacements enabled probing electromechanical phenomena in a broad range of materials ranging from ferroelectrics to ionic systems to biosystems. The main body of this work has been centered on classical ferroelectric materials, in which observed electromechanical signals offered straightforward interpretation in terms of the ferroelectric domain structures and dynamics. Combined with relatively large (10-100 pm/V) piezoelectric constants, this made PFM the leading technique for almost two decades and enabled the rapid development of this field. An important component of this progress was the emergence of effective phase-field models that allowed simulation of ferroelectric domain dynamics as a function of global (temperature, pressure)[171-173, 394] and local (tip bias, tip pressure)[164, 172, 358, 359, 395] parameters and hence provide insight into local switching mechanisms.[306, 396, 397] Recent emergence of phase field models for combined ferroelectric-semiconductor and ferroelectric-surface electrochemical behaviors offers numerous future opportunities.[398, 399]

However, the progress in resonance-enhanced PFM modes and introduction of low-noise platforms by commercial vendors enabled the PFM measurements in systems with much weaker electromechanical responses. Combined with the theoretical developments predicting a broad set of electromechanical and ferroelectric phenomena on the nanoscale, this led to a broad gamut of studies exploring ferroelectric-like behavior in dimensionally confined ferroelectrics, non-ferroelectric oxides, and macromolecular and biological systems. Here, we provide an overview of these observation, and summarize possible alternative mechanisms leading to such behaviors including both material response and electrostatic interactions, and delineate some strategies to differentiate these.

We present a (potentially incomplete) summary of alternate origins for the electromechanical displacement – a displacement that is often interpreted as stemming from a piezoelectric response. Further, we explore the methods through which alternate mechanisms can be distinguished from piezoelectricity. We show that electromechanical hysteresis does not imply ferroelectric behavior, given that application of voltage can lead to a wide variety behavior including Joule heating, charge injection and ionic motion, that can all lead to the hysteretic behavior. More advanced AFM techniques such as contact Kelvin probe force microscopy, which can be used to establish the presence of two switchable polarization states, as well as frequency-dependent studies (and studies of higher harmonics) of the electromechanical response are discussed. These methods can help to aid in the differentiation between the typical ferroelectrics from non-ferroelectrics through AFM measurements. It should be noted that the behavior of canonical relaxors is a unique case, wherein the lack of switchable polarization states dictates that they (at present) cannot be distinguished from the class of 'strange ferroelectrics' through the nanoscale measurements.

Generally, we believe that the future of this field is very bright. Despite the intrinsic complexity of coupled ferroelectric-electrochemical or ionic-electromechanical phenomena, these coupled mechanisms are now emerging at the forefront of modern science both in macro- and microscopic systems. Voltage modulated SPMs are powerful quantitative tools that allow exploration of these in real space with nanometer resolution, map in space, and correlate with specific microstructural features. As such, development of corresponding theoretical models either based on numerical schemes or approximate analytical solutions is a high priority. Similarly, of interest is development of advanced calibration and detection schemes in SPM that will allow experimental measurements of frequency dependent responses, separating frequency dispersion of materials behavior from cantilever dynamics. Synergistically, these advances can make voltage modulated SPM the local analogs of established methods such as impedance spectroscopy.

We further note that progress in the field requires development of multimodal imaging techniques combining PFM detection and structural sensitivity (e.g. focused X-ray methods[400, 401]), and chemical sensitivity (e.g. Raman or secondary ion mass-spectrometry[299, 402, 403]). These combinations can provide direct insight into physical and chemical phenomena induced by SPM tip locally. Recent studies by Ievelev *et al.*[299] illustrating non-trivial changes in surface chemistry after PFM experiments illustrate some of the serendipitous discoveries possible in this direction. Finally, synergy of SPM and high resolution STEM can reveal associated mechanisms of the atomic level. [209, 379, 404-406]

Further broad opportunities are opened by implementation of these techniques in liquid environments.[329, 407-410] While in this case electromechanical responses of material will compete with that of polarizable liquids[411, 412], these studies will provide insight into electromchanical couplings in biological systems and open the window into electro mechanics of molecular systems, the crucial development that will allow not only "to think" but also "to act" on the nanoscale.

**Acknowledgements**

We acknowledge the discussions, references to previous works, collaborations, support and friendship of all our colleagues in the ferroelectrics and scanning probe communities, without whom the progress indicated in this review would never have eventuated. We would also like to specifically thank Jan Petzelt, Stanislav Kamba, Nagarajan Valanoor, Lane Martin, Sidney Cohen, Glen Fox, Stephen Ducharme, Jim Scott, Nicola Spaldin, Brian Rodriguez and Alexei Gruverman.

A portion of this research was sponsored by the Division of Materials Sciences and Engineering, BES, DOE (RKV, SVK, PM). This research was conducted and partially supported (SJ, NB) at the Center for Nanophase Materials Sciences, which is a US DOE Office of Science User Facility.

**TABLE I: Strange Ferroelectrics in the Literature**

| Material System | Observations | | | | | | Reported Mechanisms | References |
|---|---|---|---|---|---|---|---|---|
| | Electromechanical Response | Unsaturated Hysteresis Loops | Bias Writing | Pressure Writing | Relaxation (not via wall motion) | Continuous states | | |
| Porcine aortic walls | Yes | Yes | Yes | | | | Intrinsic biological ferroelectricity | Li et al., *PRL* **108**, 078103 (2012). Ref [413] |
| $CaTiO_3$ on (001) $LaAlO_3$ with $SrRuO_3$ electrodes | Yes | | Yes | | | | Localized defect states (defect dipoles) resulting in ferroelectricity | Yang et al., *Current Applied Physics* **14**, 757 (2014). Ref [414] |
| Ultra thin $CuInP_2S_6$ flakes | Yes | | Yes | | | | Intrinsic ferroelectric state remaining in ultra-thin flakes. | Liu et al., *Nat. Comm.* **7**, 12357 (2016). Ref [296] |
| Elastin | Yes | Yes | | | | | Intrinsic polarization of individual monomer | Liu et al., *PNAS* **111**, E2780 (2014). Ref [415] |
| diphenylalanine peptide nanotubes | Yes | | Yes | | | | Co-operative proton tautomerism based on ordering and stacking of sheets with N-H..O hydrogen bonding | Bdikin et al., *Appl. Phys. Lett.* **100**, 043702 (2012). Ref [416] |
| $Sb_2S_3$ nanowires | | Yes | | | | | Orthorhombic structure, displacement of Sb and S along the c-axis. | Varghese et al., *Nano Lett.* **12**, 868 (2012). Ref [417] |
| Strain-free $SrTiO_3$ thin films | Yes | Yes | Yes | | Yes | | Sr deficiency leading to relaxor behavior | Jang et al., *PRL* **104**, 197601 (2010). Ref [418] |

| Material | Col2 | Col3 | Col4 | Col5 | Col6 | Col7 | Mechanism | Reference |
|---|---|---|---|---|---|---|---|---|
| GaFeO$_3$ thin film | Yes | | Yes | | | | Fe-driven ferroelectricity | Mukherjee et al., *PRL* **111**, 087601 (2013). Ref[419] |
| Crystalline γ-Glycine | Yes | | Yes | | | | Polar nature of individual glycine molecule with local self-assembly | Heredia et al., *Adv. Funct. Mater.*, **22**, 2996 (2012). Ref[420] |
| TbMnO$_3$ thin film (hexagonal) | Yes | Yes | Yes | Yes | | Yes | MnO5 tilt/rotation inducing polarization in unit cell | Kim et al., *Adv. Mater.*, **26**, 7660 (2014). Ref[421] |
| Nanocrystalline Hydroxyapatite Thin Films on Silicon | Yes | | Yes | | | | Ferroelectric ordering of the OH- ions | Lang et al., *Sci. Rep.* **3**, 2215 (2013). Ref[422] |
| La$_2$Zr$_2$O$_7$ thin films | Yes | Yes | Yes | | | | Non-centro symmetric structure induced in thin film geometry | Saitzek et al., *Mater. Chem. C*, **2** 4037 (2014). Ref[423] |
| LaAlO$_3$-SrTiO$_3$ Thin Film | Yes | Yes | Yes | | Yes | | Electrochemical in origin, not related to ferroelectricity [Kumar et al]; Switchable dipole from oxygen vacancy motion through the LAO layer [Bark et al.] | Kumar et al., *ACS Nano* **6** 3841 (2012). Ref[284]<br><br>Bark et al., *Nano Lett.* **12**, 1765 (2012). Ref[283] |
| h-LuFeO$_3$ | Yes | | Yes | | | | Polar structure, space group P63cm | Wang et al., *PRL* **110**, 237601 (2013). Ref[424] |
| 2 monolayer Polyvinylidene | Yes | | Yes | | | | PVDF is known ferroelectric | Yuan et al., *Materials Letters* **65**, 1989 (2011). |

| Material | | | | | | | Mechanism | Reference |
|---|---|---|---|---|---|---|---|---|
| fluoride (PVDF) with trifluoroethylene (TrFE), P(VDF-TrFE) on LaNiO$_3$/SrTiO$_3$ | | | | | | | | Ref[425] |
| SbSI nanorods | Yes | | Yes | | | | Polar chains of SBSI parallel to c-axis | Varghese et al., *Chem. Mater.* **24**, 2379 (2012). Ref[426] |
| Abalone Seashell | Yes | Yes | Yes | | | Yes | Biopolymer domains | Li and Zeng, *J. Appl. Phys.* **113**, 187202 (2013). Ref[427] |
| LaAlO$_3$ Thin films | Yes | | Yes | Yes | | Yes | Field-induced ion migration in bulk of film | Sharma et al., *Adv. Func. Mater.* **25**, 6538 (2015). Ref[216] |
| Si-doped HfO$_2$ | Yes | | Yes | | | | Intrinsic ferroelectricity in the doped compound | Martin et al. *Adv. Mater.* **26**, 8198 (2014). Ref[428] |
| Sm$_2$Ti$_2$O$_7$ Thin Film | Yes | Yes | Yes | | | | Strain-induced ferroelectricity when grown on (110) SrTiO3 | Shao et al., *J. Mater. Chem.* **22**, 9806 (2012). Ref[429] |
| SrTaO$_2$N Thin Film | Yes | Yes | Yes | | Yes | | trans-type anion ordering; strain-induced | Oka et al., *Sci. Rep.* **4** 4987 (2014). Ref[430] |
| (110) SrTiO$_3$ crystal at high temperature (420K) | Yes | | Yes | | | | Unknown, authors state not conclusive evidence for ferroelectricity. | Jyotsna et al., *J. Appl. Phys.* **116**, 104903 (2014). |
| SrTiO$_3$ Thin Films on (100) Rh | Yes | | Yes | | | | Strain-induced (tetragonal phase) | Maeng et al., *Solid State Communications* **152**, 1256 (2012). Ref[431] |
| AgSbSe$_2$ | Yes | | | | | | Lone-pair electrons distorting structure inducing local dipole | Aggarwal et al., *Appl. Phys. Lett.* **105**, 113903 (2014). Ref[432] |

| Material |  |  |  |  |  |  | Notes | Reference |
|---|---|---|---|---|---|---|---|---|
| Ultra-thin TiO$_2$ thin films on (110 NdGaO$_3$ | Yes | | Yes | | | | Initially posited as a strain-induced phenomena, later structural studies could not find symmetry breaking, discounting earlier result | Initial report: Depak et al. *Adv. Func. Mat.* **24**, 2844 (2014). Ref[344]<br><br>Subsequent Study: Skiadopoulou et al., *Adv. Func. Mat.* **26**, 642 (2016). Ref[345] |
| Strain-free SrTiO$_3$ (24UC) | Yes | | Yes | | | | Size reduction and PNR interplay | Lee et al. *Science* **349**, 1314 (2015). Ref[433] |
| Cu-doped ZnO | Yes | | Yes | | | | Unclear; Possibly related to oxygen vacancies | Xiao et al., *Acta. Mat.* **123**, 394 (2017). Ref[434] |
| BiMnO$_3$ Thin Films | Yes | | Yes | Unclear; relaxation measured over two days | | | Strain-induced distortion from ordinarily non-ferroelectric bulk structure | De Luca et al., *Appl. Phys. Lett.* **103**, 062902 (2013)[435]. |

**TABLE II: Equations for displacement from bias-induced phenomena**

| Mechanism | Displacement | Susceptibility | Time constant | Definitions |
|---|---|---|---|---|
| **Ferroelectricity** | $U = QP^2$ | $d_{eff} = \varepsilon_o Q P_{ferro}$ | $\tau_s = \tau_d = ns$ | U = Surface Displacement<br>Q = electrostrictive coefficient<br>R = Tip Radius |
| **Chemical Dipole** | " | $d_{chem} = \varepsilon_o Q P_{chem}$ | $\tau_s = R\lambda/D$, $\tau_d = ns$ | β = Vegard coefficient<br>λ = Debye length<br>$D_{ion/therm}$ = ion or thermal diffusion coefficient |
| **Charge Injection** | " | " | " | P = Polarization<br>α = thermal expansion coefficient |
| **Field Effect** | " | " | " | ρ = resistivity<br>i = current<br>ω = frequency |
| **Vegard Strain** | $U \propto \dfrac{\beta D_{ion}}{\sqrt{\omega}}$ | $d_{Veg} \propto \dfrac{\beta D_{ion}}{\sqrt{\omega}}$ | $\tau_s = \tau_d = R^2/D_{ion}$ | $\tau_s$ = time for state change<br>$\tau_d$ = time for surface deformation |
| **Joule Heating** | $U = \alpha \rho i^2$ | $d_{Therm} \propto \dfrac{\alpha D_{therm}}{\sqrt{\omega}}$ | $\tau_s = \tau_d = R^2/D_{therm}$ | |

**Figure Captions**

**Figure 1: CiteSpace analysis of papers containing the word 'Ferroelectrics'.** Only the top 3000 cited papers have been chosen in the analysis, with dates from 1985 through to 2016. Through semantic analysis of the keywords within the abstracts of the cited papers, clusters can be formed, which are plotted on a timeline view **(a)**. The keywords identifying each cluster are indicated to the right of each tier (each tier is an individual cluster within the citation network). Nodes with large rings indicate citation 'bursts' (rapid increase in number of citations after publication). **(b)** Alternate representation of (a), with clusters colored by the (mean) year of their citers (legend below). For clarity, only highly cited works are shown.

**Figure 2: Schematic of the setup for piezoresponse force microscopy**. A voltage $V$ (which can have both DC and AC components) is applied to the tip which is in contact with the sample. This produces a strong local field in the material, which can lead to volume changes, resulting in a force on the tip which is measured as a displacement of the cantilever via a beam deflection system. At the same time, the tip and cantilever will feel electrostatic forces from the potential difference between tip and sample, that can also contribute to the measured displacement.

**Figure 3: Examples of typical use cases of PFM.** (Upper Row): PFM images of domains in three different samples – a ceramic, and two thin films, showing distinct domain structures. (Lower Panel), PFM images after poling experiments to manipulate the domains for a ceramic and two thin films. In the case of the $BiFeO_3$ (BFO) thin film, specific poling procedures can be followed to create complex domain patterns as a result of the rhombohedral structure with 8 equivalent directions (along <111>).

**Figure 4: Piezoelectric response in biological samples. (a,b)** Topography and Vector PFM map of the response from proteins in human teeth. The piezoelectric response from a butterfly wing is seen in the topography and Vector PFM map in **(c,d)**. (a-d) reprinted from Kalinin *et al*.[436], with permission from Elsevier. Piezoelectric response is also seen in PFM studies of a seashell, as seen in **(e)** Topography, **(f)** Vertical PFM Amplitude and **(g)** Phase images. (e-g) are reprinted from Li *et al*.[427], with the permission of AIP Publishing

**Figure 5: Examples of spatially resolved switching in the vicinity of structural and topological defects. (a)** Ferroelectric **s**witching in the vicinity of a grain boundary in a ferroelectric thin film grown on a bi-crystal. Shown here is an imprint map, indicating the shift of the hysteresis loop along the voltage axis. **(b)** Selected hysteresis loops from points indicated in (a). Switching loops near the grain boundary (blue, green) display different characteristics, including a kink, than those away from the grain boundary. **(c)** Phase-field modeling confirms that nucleation potential is substantially affected by the presence of the grain boundary, and will lead to asymmetry, which is seen in experiment. The same method can be applied to pre-existing domain walls (topological defects), as in **(d-f)**. (d) Selected hysteresis loops acquired on domain walls and within domains, showing the decreased nucleation bias on the negative to positive

branch. This can be directly mapped, as shown in (d), and corresponds with the positions of the in-plane domain walls (labelled IP DW in the figure). (f) Phase-field modeling suggests a twisted structure developing that increases the field locally, and reduces the voltage necessary to initiate nucleation of the reverse domain. Images from (a-c) reproduced from Rodriguez *et al.*[170] and images from (d-f) reproduced from Balke *et al.*[164]

**Figure 6: Strange Ferroelectricity examples from the literature. (a-f)** Examples of ferroelectric hysteresis loops include strain-free $SrTiO_3$ (a), $LaAlO_3/SrTiO_3$ (b), $Sb_2S_3$ (c), and ZnO (d). Reports of pressure-induced switching have also been made, such as the PFM images with vertical PFM amplitude and phase, respectively in (e-h), for $BaTiO_3$ (e,f) and $LaAlO_3$ (g,h). Figure in (a) is from Jang *et al.*[418], (b) Reprinted with permission from Bark *et al.*[283], Copyright 2012 American Chemical Society. (c) is Reprinted with permission from Varghese *et al.*[417], Copyright 2012 American Chemical Society. (d) is from ref[437], (e,f) is reproduced from Liu *et al.*[223] and (g,h) is from Sharma *et al.*[216].

**Figure 7: Biological ferroelectricity**. These include examples of hysteretic switching shown as amplitude and phase of the measured PFM signal in biomaterials such as aortic walls (a-d) and elastin (e-h). (a-d) are reprinted from Liu *et al.* [413], with the permission of AIP publishing. (e-h) are reproduced from Liu *et al.*[415]

**Figure 8: Bias-induced phenomena in contact atomic force microscopy**. These can range from purely physical effects, such as ferroelectricity, to electrochemical effects such as vacancy ordering, deposition, and damage (breakdown). Alongside each mechanism are the expected changes ($\Delta$) in concentration (n), charge (q) and dipole moment (p) under an applied electric field.

**Figure 9: Differentiating the different mechanisms from tip bias-induced experiments.** Presented in this figure are some characteristic features of the various physical and (electro)chemical processes that can occur in contact AFM when bias is applied, and how they may be separated.

**Figure 10: Differentiating ferro/piezoelectric from non-ferro/piezoelectric response. (a,b)** Hysteresis loops acquired by PFM on BFO nanocapacitors, as a function of the applied AC potential. As the amplitude of AC is increased, the loop width reduces gradually, while the loop height remains mostly constant, until the applied voltage is comparable to the coercive field. At this point, the loop height collapses. Reprinted from Strelcov *et al.*[334] with the permission of AIP Publishing. Another method to determine whether a material is piezoelectric is to determine the frequency dispersion, as shown by Seol *et al.*[336] in **(c-d)**. (c) Amplitude of the electromechanical response on PZT as a function of frequency. The response is mostly invariant to the frequency. (d) Similar measurement on LICGC, showing strong dispersion with frequency of the applied $V_{ac}$. Reproduced from Seol *et al.*[336]. A similar example is indicated by Chen *et al.*[118] **(e-h)** Comparison

of soda-lime glass (e,f) and PZT sample (g,h) hysteresis loops as a function of voltage window and frequency of the applied DC switching waveform. The variance is much larger for the glass than the PZT sample, and the change in amplitude is especially pronounced, while for the PZT sample the amplitude change is minimal with respect to the time taken for the DC sweep. Reprinted from Chen *et al.*[118] with the permission of AIP Publishing.

**Figure 11: Comparison of on-field and off-field loops for ferroelectric PZT, as well as HfO₂.** **(a)** off-field measurements of the electromechanical displacement $D_{ac}$, and **(b)** on-field measurements of the same samples.

**Figure 12: Comparison of standard hysteresis loop acquisition with contact KPFM measurements.** (a) DC Waveform used for standard hysteresis measurements. (b) In cKPFM, the DC waveform is similar, but we additionally vary the voltage during the reading step after each pulse, and repeat the waveform for measuring the response at different $V_{read}$ voltages. (c) We compare remnant PFM hysteresis loops for PZT on $La_{1-x}Sr_xMnO_3$ (LSMO) electrode, non-ferroelectric $SrTiO_3$ on nearly lattice-matched $DyScO_3$ and an amorphous HfO₂ thin film. All the loops qualitatively resemble ferroelectric behavior, which is expected for PZT, may occur for strained $SrTiO_3$ and should not be observed in $SrTiO_3/DyScO_3$ substrates. Shown in **(d)** are the results of contact KPFM measurements on the same three samples. The acquired loops from cKPFM measurement are plotted as a function of the read DC voltage. Figure is reproduced (adapted) from Balke *et al.*[335]

**Figure 13: Measuring the junction potential and capacitance gradient.** Shown in **(a)** is an example of a cKPFM measurement on a dielectric (non-ferroelectric) sample. The *x*-intercepts of each line show the junction potential **(b)**, which displays hysteresis. At the same time, the results taken across the y-axis (at $V_{read} = 0V$) are the same as those that would be obtained from a standard off-field PFM hysteresis measurement, which is shown in **(c)**. The slopes of the lines, which is proportional to the capacitance gradient *C'* (and which is a measure of the sensitivity of the cantilever to the electrostatic force) can also be plotted, as in **(d)**. This shows that at different values of surface potentials, the value of *C'* is not constant, but rather indicates a peak in this case near ~0.5V. Figure is adapted from Balke *et al.*[335]

**Figure 14: cKPFM measurement on a HfO₂ thin film**. **(a)** Standard KPFM measurement of the surface of HfO₂ thin film after two areas were 'poled' by applying DC bias to the tip and scanning the sample in contact-mode. This produces two 'domains' as marked by the bright and dark contrasts in the KPFM image of the surface potential. **(b)** Measurement of the surface potential by applying the cKPFM technique. The surface potential can be measured using cKPFM mode by determining the x-intercept of the lines in the cKPFM measurement, in the absence of spontaneous polarization states, i.e., by looking at the intercept of each individual line in Fig. 13a, one can obtain the 'cKPFM' surface potential. The line profile comparison over the two domains from the standard non-contact KPFM, and cKPFM, is shown in **(c)**. 'PFM' amplitude and phase images of the same domains are shown in **(d,e)**. For comparison, PFM images of amplitude and phase from a poled ferroelectric PZT thin film are shown in **(f,g)**. (a-c) is reproduced from Balke *et al*. [346]

**Figure 15**: **Quantification of electrostatics.** A procedure to enable quantitative measurements independent of the chosen cantilever is detailed in Balke *et al*. [349] This method was used to determine the effective $d_{33}$ values of a LiNbO$_3$ single crystal, but requires quantitative determination of the electrostatic and piezoelectric contributions to the response. The amplitude of the displacement $D_{ac}$ for two different cantilevers A and B is shown in **(a)** for two differently oriented domains. Clearly, the slopes of the response differ due to the different contact stiffness values of the cantilevers. After using a correction factor accounting for individual cantilever geometry (so-called shape factor λ), it becomes possible to compute the displacement directly, as in **(b)**. The results, without using this calibration are plotted in the upper panel, and after accounting for this factor, are shown in the lower panel. Shown in **(c,d)** are separation of the signal $D_{ac}$ into **(c)** electrostatic, $D_{ES}$, and **(d)** piezoelectric, $D_{PE}$, contributions plotted as functions of the measured PPLN surface potential $V_{SP}$ for cantilevers A and B. For comparison, panel (d) also shows the value of the piezoelectric constant specified by the periodically poled lithium niobate crystal provider. Figures reproduced from Balke *et al*.[349]

**Figure 16: Acoustic Detection of field-induced phase transitions in ferroelectrics. (a)** Piezoresponse hysteresis loop and **(b)** Resonant frequency acquired simultaneously from PFM spectroscopy of a PMN-0.28PT single crystal. These fields are sufficient to cause local rhombohedral to tetragonal (or monoclinic) phase transformations, with characteristic softening (see schematic of process in **(c)**). Reproduced from Vasudevan *et al.*[358] Similar experiments on a BiFeO$_3$ thin films are shown in **(d)**. Here, the piezoresponse is plotted as a function of negative bias applied to the PFM tip, along with the resonant frequency. On-field loops are plotted in red/blue and off-field loops are in black/cyan. In this case, the rhombohedral-tetragonal phase transformation (magenta background region) is only seen in the on-field loops, where softening is clearly observed. **(e,f)** Phase-field simulations of the in-plane switching and rhombohedral-tetragonal phase transition by application of tip bias. A small volume under the tip is transformed in the on-field state. Images reproduced from Li *et al.*[359]

**Figures**

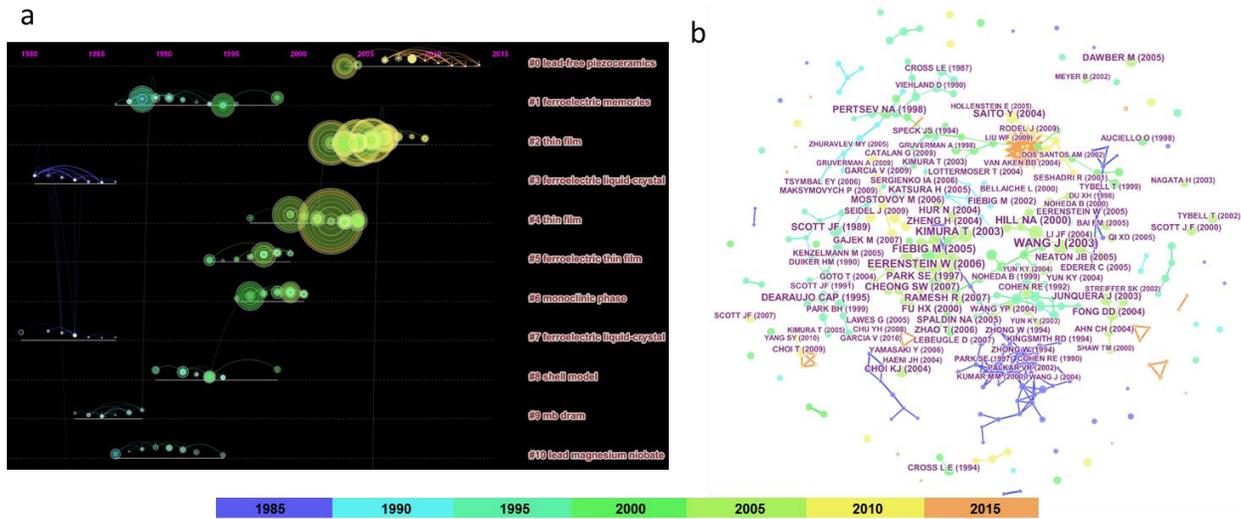

**Figure 1: CiteSpace analysis of papers containing the word 'Ferroelectrics'.** Only the top 3000 cited papers have been chosen in the analysis, with dates from 1985 through to 2016. Through semantic analysis of the keywords within the abstracts of the cited papers, clusters can be formed, which are plotted on a timeline view **(a)**. The keywords identifying each cluster are indicated to the right of each tier (each tier is an individual cluster within the citation network). Nodes with large rings indicate citation 'bursts' (rapid increase in number of citations after publication). **(b)** Alternate representation of (a), with clusters colored by the (mean) year of their citers (legend below). For clarity, only highly cited works are shown.

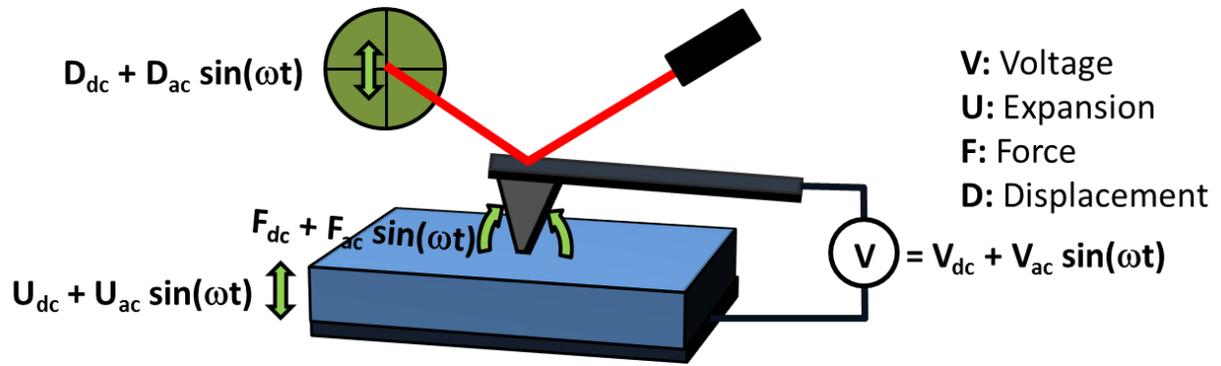

**Figure 2: Schematic of the setup for piezoresponse force microscopy**. A voltage *V* (which can have both DC and AC components) is applied to the tip which is in contact with the sample. This produces a strong local field in the material, which can lead to volume changes, resulting in a force on the tip which is measured as a displacement of the cantilever via a beam deflection system. At the same time, the tip and cantilever will feel electrostatic forces from the potential difference between tip and sample, that can also contribute to the measured displacement.

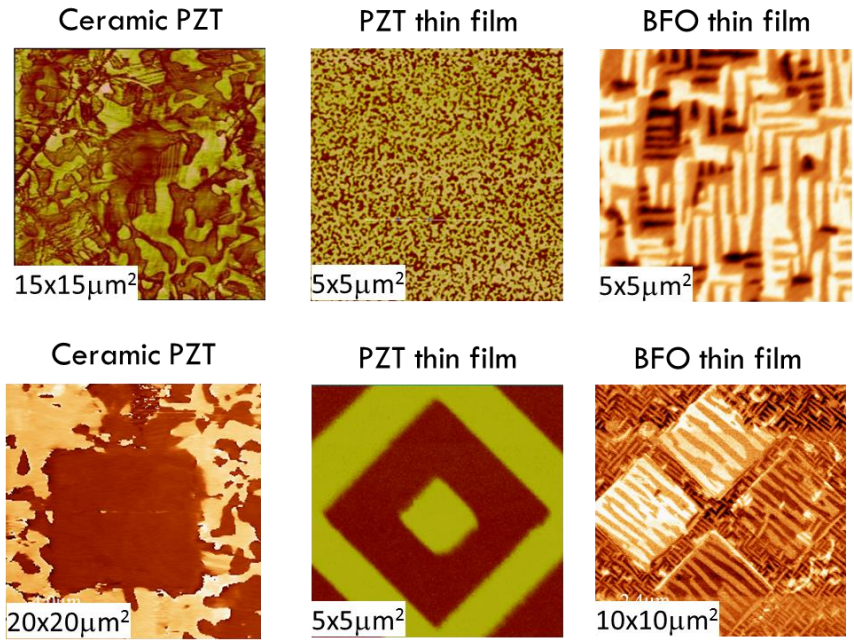

**Figure 3: Examples of typical use cases of PFM.** (Upper Row): PFM images of domains in three different samples – a ceramic, and two thin films, showing distinct domain structures. (Lower Panel), PFM images after poling experiments to manipulate the domains for a ceramic and two thin films. In the case of the BiFeO$_3$ (BFO) thin film, specific poling procedures can be followed to create complex domain patterns as a result of the rhombohedral structure with 8 equivalent directions (along <111>).

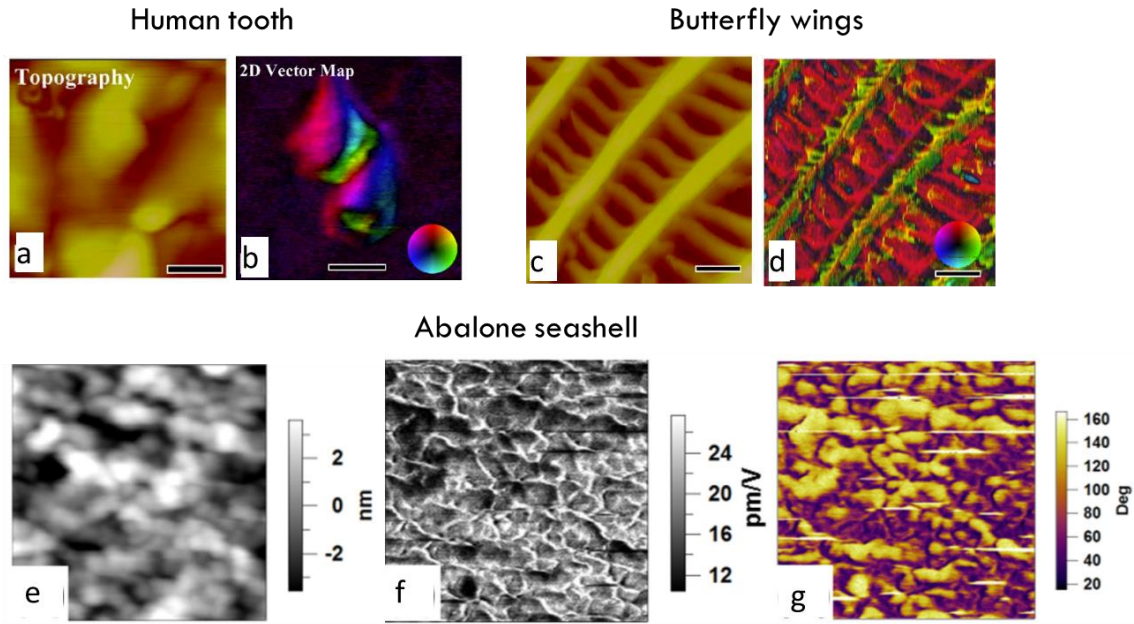

**Figure 4: Piezoelectric response in biological samples. (a,b)** Topography and Vector PFM map of the response from proteins in human teeth. The piezoelectric response from a butterfly wing is seen in the topography and Vector PFM map in **(c,d)**. (a-d) reprinted from Kalinin *et al.*[436], with permission from Elsevier. Piezoelectric response is also seen in PFM studies of a seashell, as seen in **(e)** Topography, **(f)** Vertical PFM Amplitude and **(g)** Phase images. (e-g) are reprinted from Li *et al.*[427], with the permission of AIP Publishing

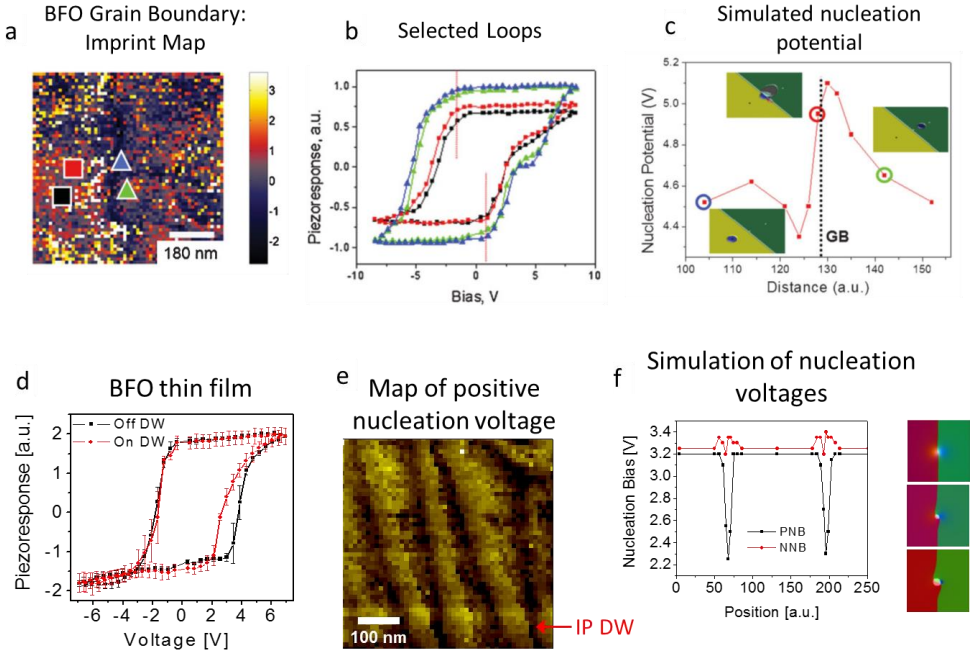

**Figure 5: Examples of spatially resolved switching in the vicinity of structural and topological defects. (a) Ferroelectric s**witching in the vicinity of a grain boundary in a ferroelectric thin film grown on a bi-crystal. Shown here is an imprint map, indicating the shift of the hysteresis loop along the voltage axis. **(b)** Selected hysteresis loops from points indicated in (a). Switching loops near the grain boundary (blue, green) display different characteristics, including a kink, than those away from the grain boundary. **(c)** Phase-field modeling confirms that nucleation potential is substantially affected by the presence of the grain boundary, and will lead to asymmetry, which is seen in experiment. The same method can be applied to pre-existing domain walls (topological defects), as in **(d-f)**. (d) Selected hysteresis loops acquired on domain walls and within domains, showing the decreased nucleation bias on the negative to positive branch. This can be directly mapped, as shown in (d), and corresponds with the positions of the in-plane domain walls (labelled IP DW in the figure). (f) Phase-field modeling suggests a twisted structure developing that increases the field locally, and reduces the voltage necessary to initiate nucleation of the reverse domain. Images from (a-c) reproduced from Rodriguez *et al.*[170] and images from (d-f) reproduced from Balke *et al.*[164]

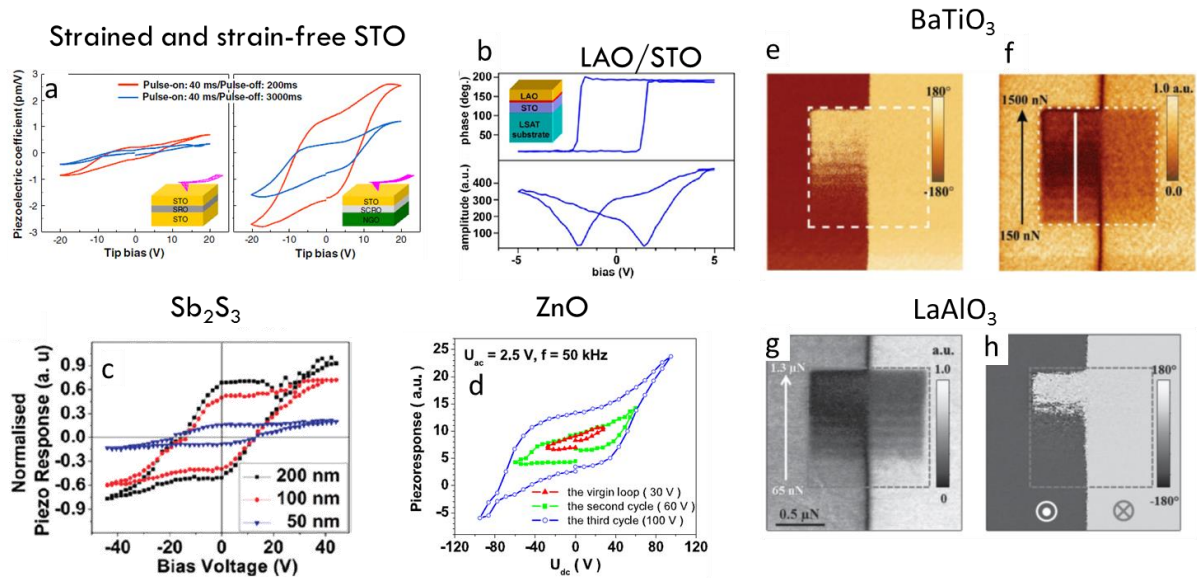

**Figure 6: Strange Ferroelectricity examples from the literature. (a-f)** Examples of ferroelectric hysteresis loops include strain-free SrTiO$_3$ (a), LaAlO$_3$/SrTiO$_3$ (b), Sb$_2$S$_3$ (c), and ZnO (d). Reports of pressure-induced switching have also been made, such as the PFM images with vertical PFM amplitude and phase, respectively in (e-h), for BaTiO$_3$ (e,f) and LaAlO$_3$ (g,h). Figure in (a) is from Jang *et al.*[418], (b) Reprinted with permission from Bark *et al.*[283], Copyright 2012 American Chemical Society. (c) is Reprinted with permission from Varghese *et al.*[417], Copyright 2012 American Chemical Society. (d) is from ref[437], (e,f) is reproduced from Liu *et al.*[223] and (g,h) is from Sharma *et al.*[216].

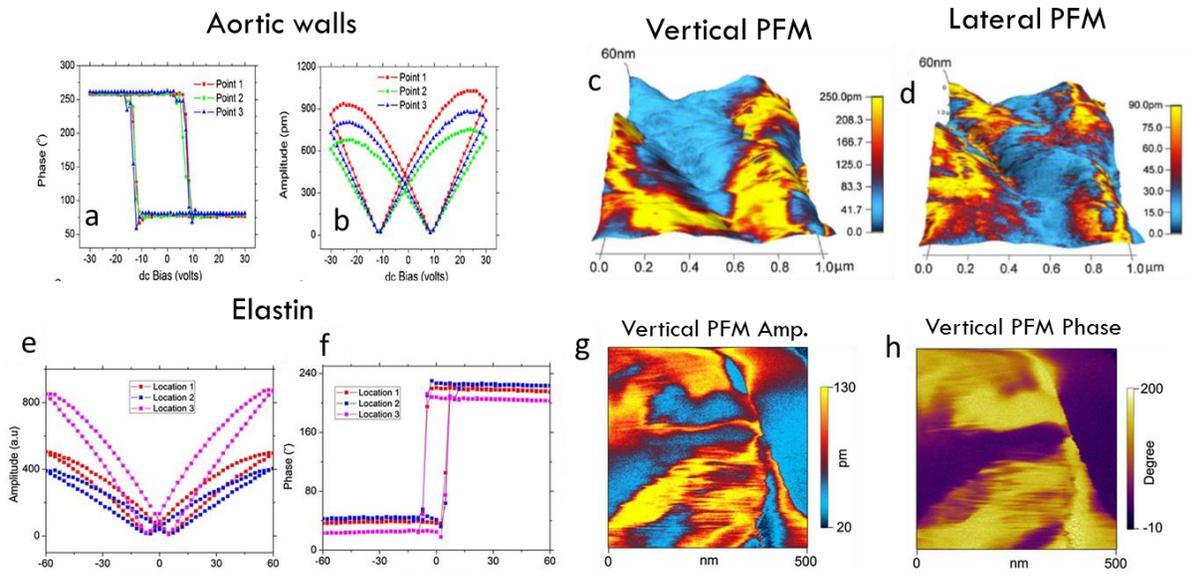

**Figure 7: Biological ferroelectricity**. These include examples of hysteretic switching shown as amplitude and phase of the measured PFM signal in biomaterials such as aortic walls (a-d) and elastin (e-h). (a-d) are reprinted from Liu *et al*.[413], with the permission of AIP publishing. (e-h) are reproduced from Liu *et al*.[415]

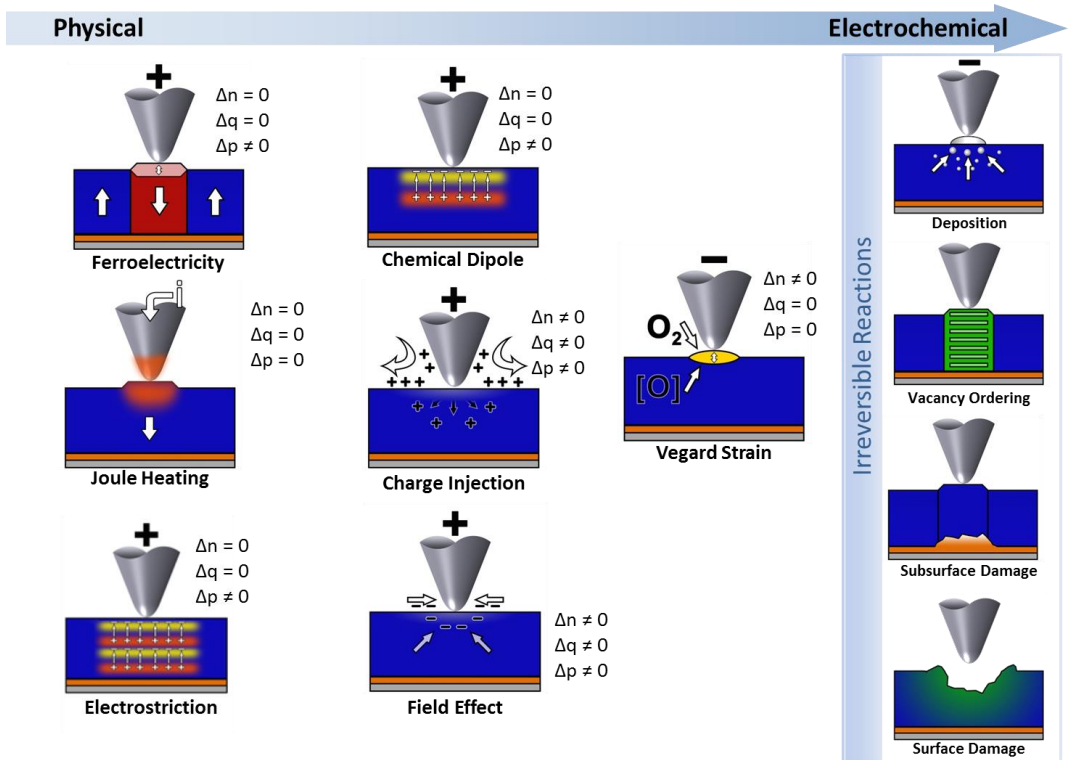

**Figure 8: Bias-induced phenomena in contact atomic force microscopy**. These can range from purely physical effects, such as ferroelectricity, to electrochemical effects such as vacancy ordering, deposition, and damage (breakdown). Alongside each mechanism are the expected changes ($\Delta$) in concentration (n), charge (q) and dipole moment (p) under an applied electric field.

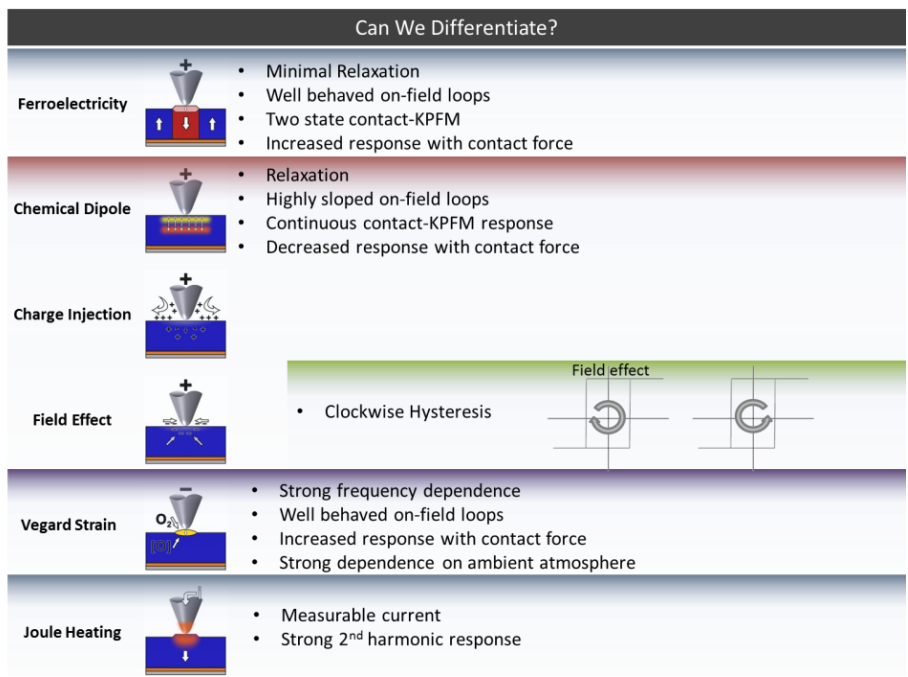

**Figure 9: Differentiating the different mechanisms from tip bias-induced experiments.** Presented in this figure are some characteristic features of the various physical and (electro)chemical processes that can occur in contact AFM when bias is applied, and how they may be separated.

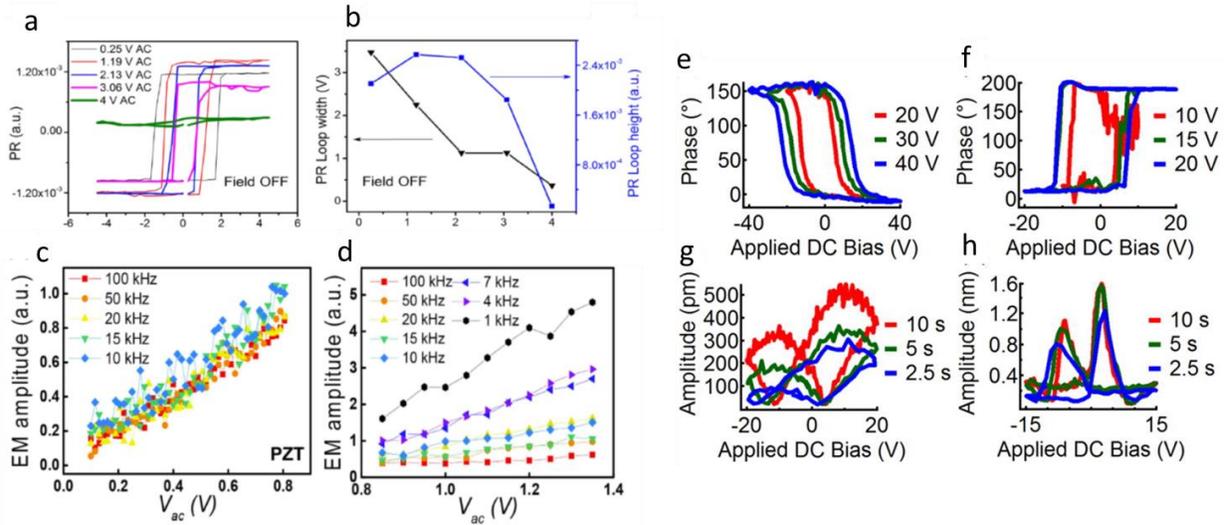

**Figure 10: Differentiating ferro/piezoelectric from non-ferro/piezoelectric response. (a,b)** Hysteresis loops acquired by PFM on BFO nanocapacitors, as a function of the applied AC potential. As the amplitude of AC is increased, the loop width reduces gradually, while the loop height remains mostly constant, until the applied voltage is comparable to the coercive field. At this point, the loop height collapses. Reprinted from Strelcov *et al.*[334] with the permission of AIP Publishing. Another method to determine whether a material is piezoelectric is to determine the frequency dispersion, as shown by Seol *et al.*[336] in **(c-d)**. (c) Amplitude of the electromechanical response on PZT as a function of frequency. The response is mostly invariant to the frequency. (d) Similar measurement on LICGC, showing strong dispersion with frequency of the applied $V_{ac}$. Reproduced from Seol *et al.*[336]. A similar example is indicated by Chen *et al.*[118] **(e-h)** Comparison of soda-lime glass (e,f) and PZT sample (g,h) hysteresis loops as a function of voltage window and frequency of the applied DC switching waveform. The variance is much larger for the glass than the PZT sample, and the change in amplitude is especially pronounced, while for the PZT sample the amplitude change is minimal with respect to the time taken for the DC sweep. Reprinted from Chen *et al.*[118] with the permission of AIP Publishing.

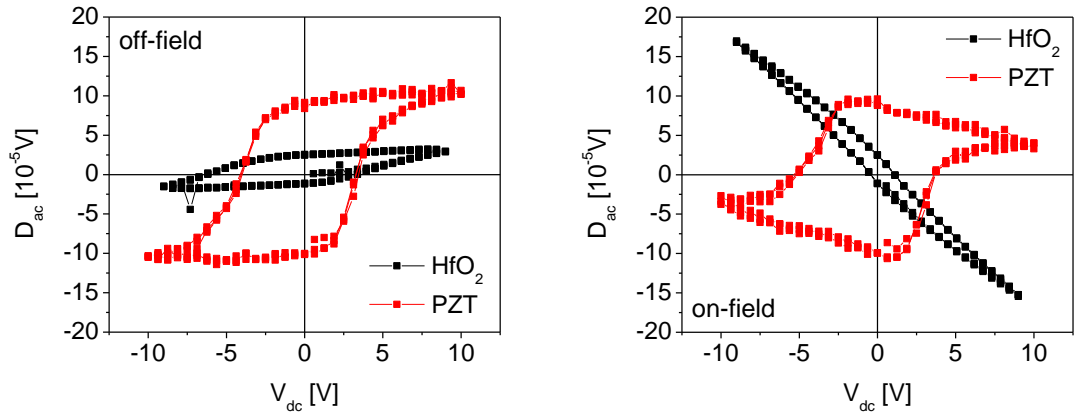

**Figure 11: Comparison of on-field and off-field loops for ferroelectric PZT, as well as HfO₂.** **(a)** off-field measurements of the electromechanical displacement $D_{ac}$, and **(b)** on-field measurements of the same samples.

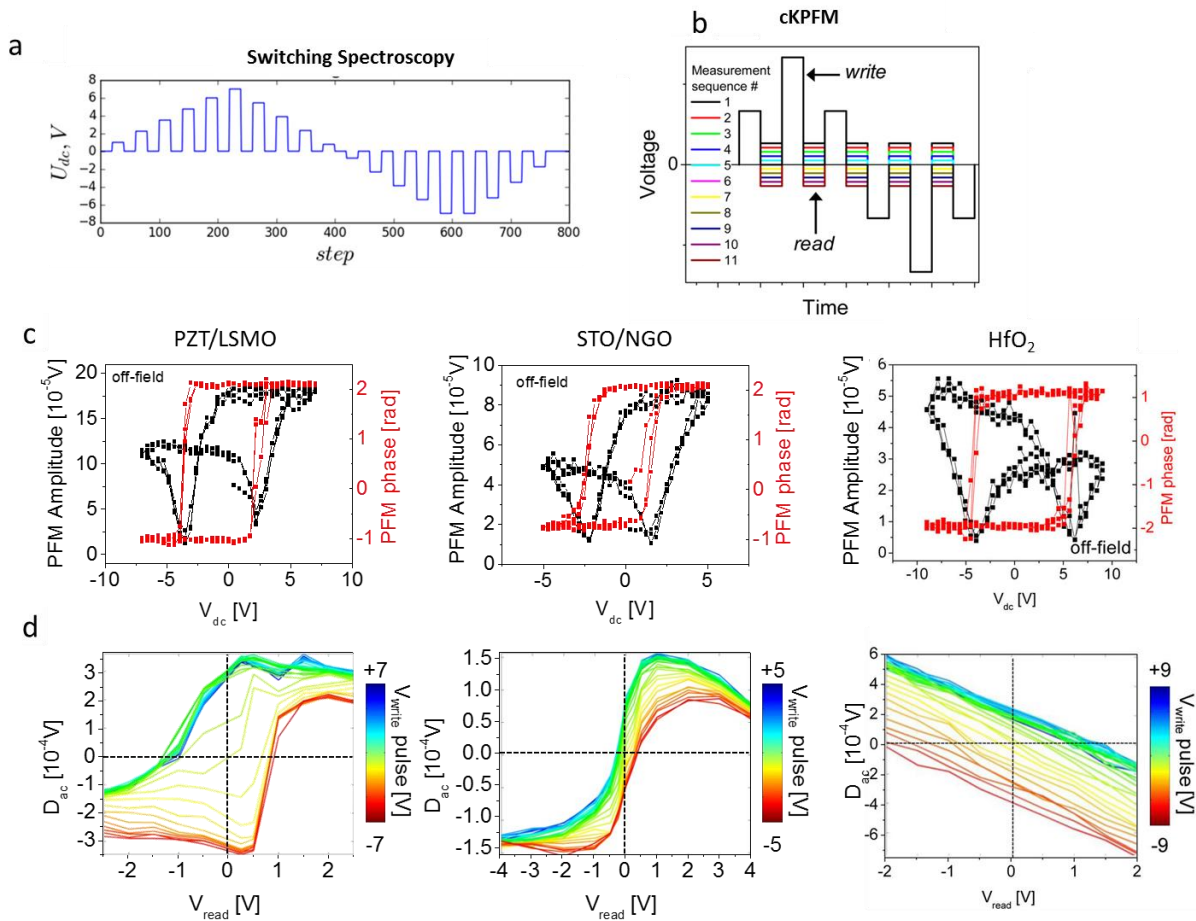

**Figure 12: Comparison of standard hysteresis loop acquisition with contact KPFM measurements.** (a) DC Waveform used for standard hysteresis measurements. (b) In cKPFM, the DC waveform is similar, but we additionally vary the voltage during the reading step after each pulse, and repeat the waveform for measuring the response at different $V_{read}$ voltages. (c) We compare remnant PFM hysteresis loops for PZT on $La_{1-x}Sr_xMnO_3$ (LSMO) electrode, non-ferroelectric $SrTiO_3$ on nearly lattice-matched $DyScO_3$ and an amorphous $HfO_2$ thin film. All the loops qualitatively resemble ferroelectric behavior, which is expected for PZT, may occur for strained $SrTiO_3$ and should not be observed in $SrTiO_3/DyScO_3$ substrates. Shown in **(d)** are the results of contact KPFM measurements on the same three samples. The acquired loops from cKPFM measurement are plotted as a function of the read DC voltage. Figure is reproduced (adapted) from Balke *et al.*[335]

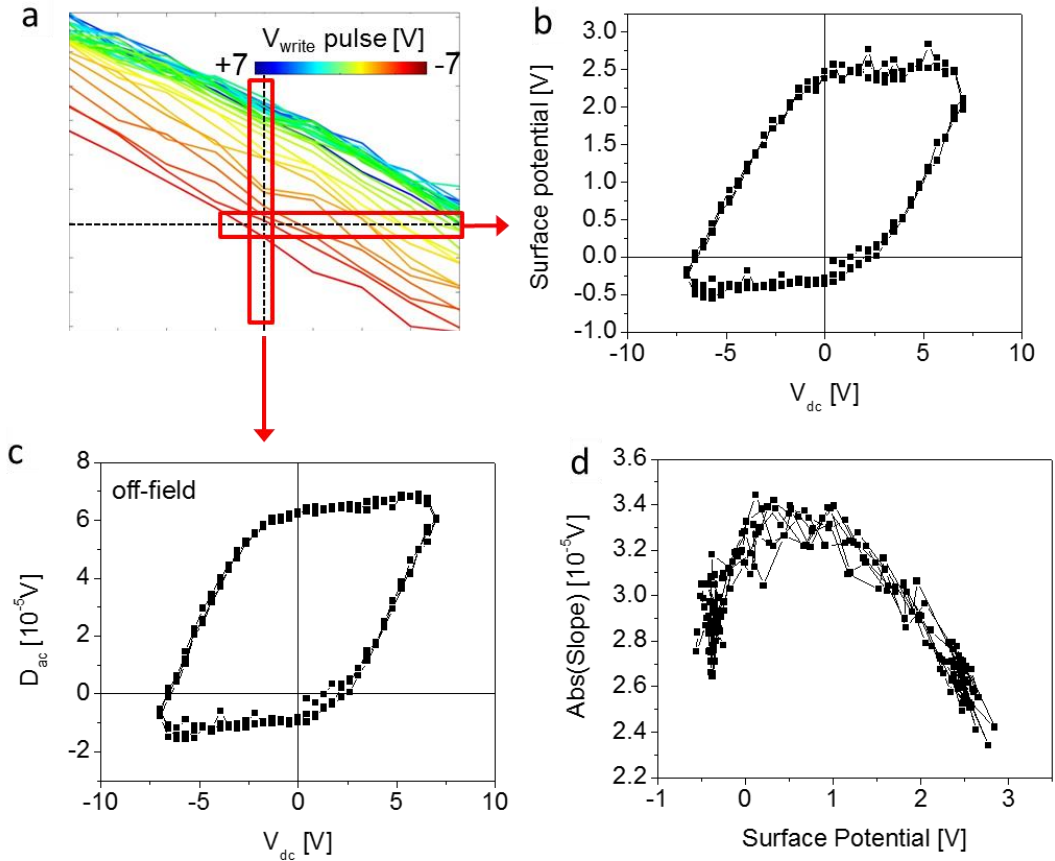

**Figure 13: Measuring the junction potential and capacitance gradient.** Shown in **(a)** is an example of a cKPFM measurement on a dielectric (non-ferroelectric) sample. The x-intercepts of each line show the junction potential **(b)**, which displays hysteresis. At the same time, the results taken across the y-axis (at $V_{read} = 0V$) are the same as those that would be obtained from a standard off-field PFM hysteresis measurement, which is shown in **(c)**. The slopes of the lines, which is proportional to the capacitance gradient *C'* (and which is a measure of the sensitivity of the cantilever to the electrostatic force) can also be plotted, as in **(d)**. This shows that at different values of surface potentials, the value of *C'* is not constant, but rather indicates a peak in this case near ~0.5V. Figure is adapted from Balke *et al*.[335]

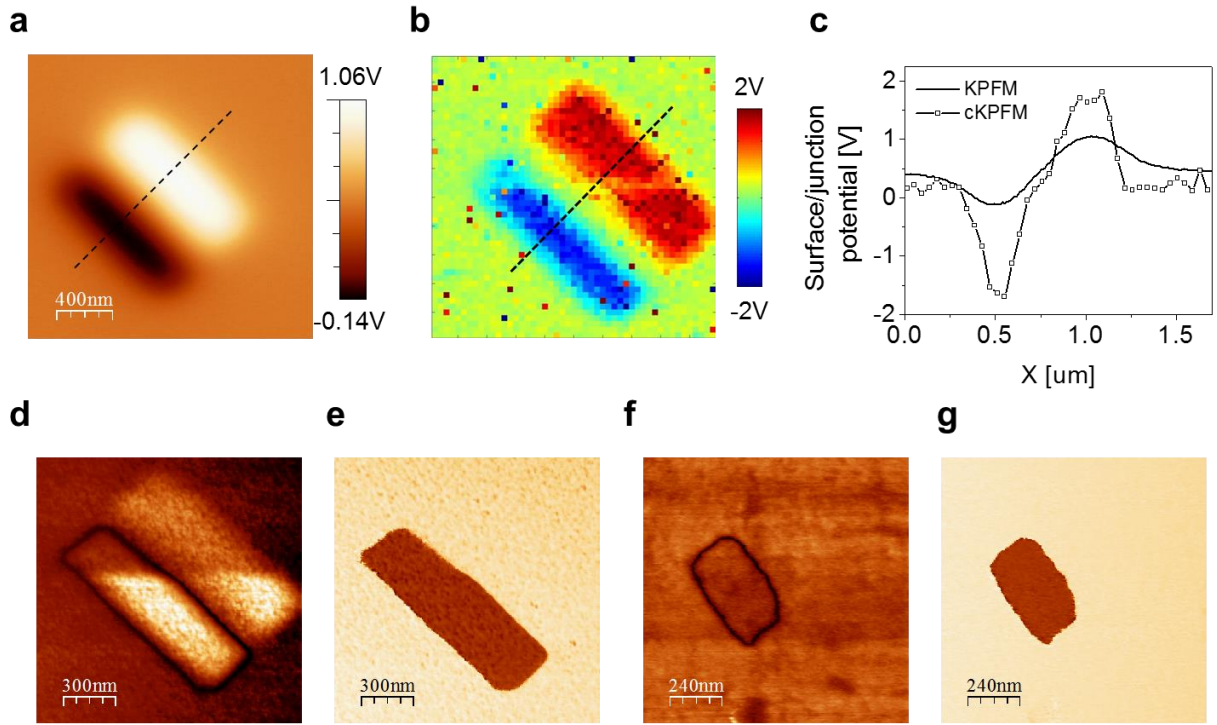

**Figure 14: cKPFM measurement on a HfO$_2$ thin film**. **(a)** Standard KPFM measurement of the surface of HfO$_2$ thin film after two areas were 'poled' by applying DC bias to the tip and scanning the sample in contact-mode. This produces two 'domains' as marked by the bright and dark contrasts in the KPFM image of the surface potential. **(b)** Measurement of the surface potential by applying the cKPFM technique. The surface potential can be measured using cKPFM mode by determining the x-intercept of the lines in the cKPFM measurement, in the absence of spontaneous polarization states, i.e., by looking at the intercept of each individual line in Fig. 13a, one can obtain the 'cKPFM' surface potential. The line profile comparison over the two domains from the standard non-contact KPFM, and cKPFM, is shown in **(c)**. 'PFM' amplitude and phase images of the same domains are shown in **(d,e)**. For comparison, PFM images of amplitude and phase from a poled ferroelectric PZT thin film are shown in **(f,g)**. (a-c) is reproduced from Balke *et al*.[346]

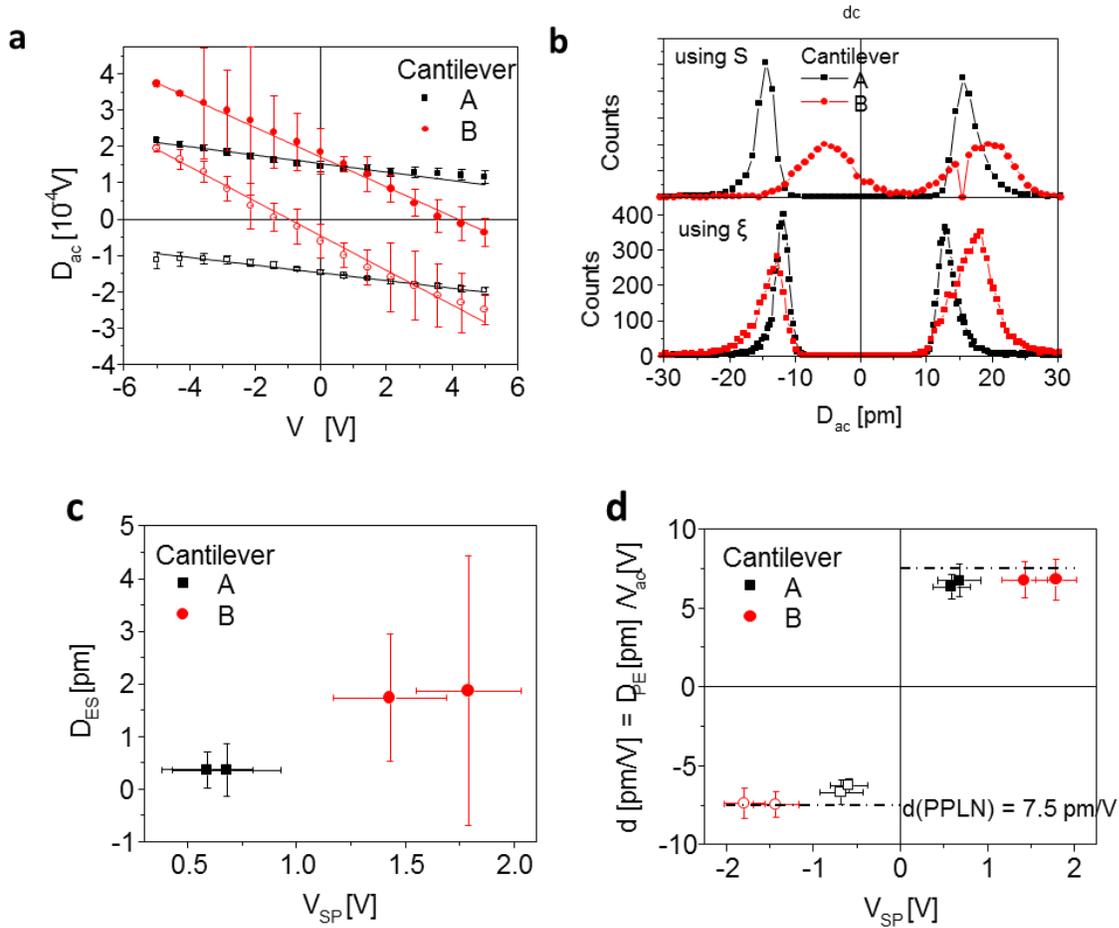

**Figure 15**: **Quantification of electrostatics.** A procedure to enable quantitative measurements independent of the chosen cantilever is detailed in Balke *et al*. [349] This method was used to determine the effective $d_{33}$ values of a LiNbO$_3$ single crystal, but requires quantitative determination of the electrostatic and piezoelectric contributions to the response. The amplitude of the displacement $D_{ac}$ for two different cantilevers A and B is shown in **(a)** for two differently oriented domains. Clearly, the slopes of the response differ due to the different contact stiffness values of the cantilevers. After using a correction factor accounting for individual cantilever geometry (so-called shape factor λ), it becomes possible to compute the displacement directly, as in **(b)**. The results, without using this calibration are plotted in the upper panel, and after accounting for this factor, are shown in the lower panel. Shown in **(c,d)** are separation of the signal $D_{ac}$ into **(c)** electrostatic, $D_{ES,}$ and **(d)** piezoelectric, $D_{PE}$, contributions plotted as functions of the measured PPLN surface potential $V_{SP}$ for cantilevers A and B. For comparison, panel (d) also shows the value of the piezoelectric constant specified by the periodically poled lithium niobate crystal provider. Figures reproduced from Balke *et al*.[349]

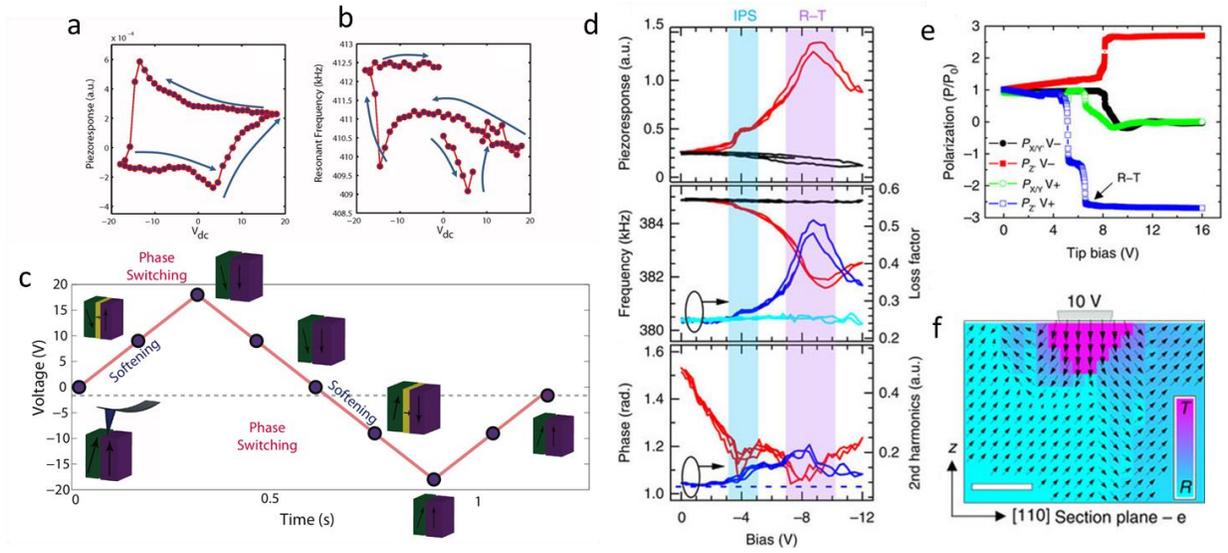

**Figure 16: Acoustic Detection of field-induced phase transitions in ferroelectrics. (a)** Piezoresponse hysteresis loop and **(b)** Resonant frequency acquired simultaneously from PFM spectroscopy of a PMN-0.28PT single crystal. These fields are sufficient to cause local rhombohedral to tetragonal (or monoclinic) phase transformations, with characteristic softening (see schematic of process in **(c)**). Reproduced from Vasudevan *et al.*[358] Similar experiments on a BiFeO$_3$ thin films are shown in **(d)**. Here, the piezoresponse is plotted as a function of negative bias applied to the PFM tip, along with the resonant frequency. On-field loops are plotted in red/blue and off-field loops are in black/cyan. In this case, the rhombohedral-tetragonal phase transformation (magenta background region) is only seen in the on-field loops, where softening is clearly observed. **(e,f)** Phase-field simulations of the in-plane switching and rhombohedral-tetragonal phase transition by application of tip bias. A small volume under the tip is transformed in the on-field state. Images reproduced from Li *et al.*[359]